\documentclass[12pt]{article}

\usepackage{amsfonts,amsbsy}
\input{epsf}
\usepackage{epsfig}

\def\IC{\mathbb{C}}

\def\IZ{\mathbb{Z}}
\def\IR{\mathbb{R}}
\def\IP{\mathbb{P}}
\def\IF{{\cal F}}
\def\sF{{\boldsymbol{\cal F}}}
\def\sG{{\boldsymbol{\cal G}}}
\def\IA{{\cal A}}
\def\sA{{\boldsymbol{\cal A}}}
\def\s*{\boldsymbol{*}}
\def\sN{{\bf N}}
\def\sd{{\bf d}}
\def\sD{{\bf D}}
\def\sx{{\bf x}}

\def\sQ{{\bf Q}}
\def\sJ{{\bf J}}
\def\se{{\bf e}}
\def\hn{\boldsymbol{|}}
\def\szeta{{\boldsymbol{\zeta}}}

\def\CN {{\cal N}}

\def\CK{{\cal K}}

\newcommand{\sect}[1]{\setcounter{equation}{0}\section{#1}}

\renewcommand{\Im}{{\rm Im}\,}
\renewcommand{\Re}{{\rm Re}\,}
\newcommand{\NP}[1]{Nucl.\ Phys.\ {\bf #1}}
\newcommand{\CQG}[1]{Class.\ Quant.\ Grav.\ {\bf #1}}
\newcommand{\PL}[1]{Phys.\ Lett.\ {\bf #1}}

\newcommand{\PR}[1]{Phys.\ Rev.\ {\bf #1}}
\newcommand{\PRL}[1]{Phys.\ Rev.\ Lett.\ {\bf #1}}

\def\one{{\hbox{ 1\kern-.8mm l}}}
\def\ii{i}
\def\bz{{\bar z}}

\newcommand{\N}{{\cal N}}

\def\bb{\bar{b}}
\def\bz{\bar{z}}
\def\bZ{\bar{Z}}
\def\bOmega{\bar{\Omega}}
\def\bpartial{\bar{\partial}}

\def\tomega{\tilde{\omega}}
\def\dop{d}

\newlength{\bredde}
\def\slash#1{\settowidth{\bredde}{$#1$}\ifmmode\,\raisebox{.15ex}{/}
\hspace*{-\bredde} #1\else$\,\raisebox{.15ex}{/}\hspace*{-\bredde} #1$\fi}

\begin{document}

\begin{titlepage}
\begin{flushright} hep-th/0005049
\end{flushright}

\vfill
\begin{center}
{\LARGE\bf Supergravity flows and D-brane stability}
\\
\vskip 20mm {\bf Frederik Denef} \\ \vskip 7mm {\em Department of
Mathematics
\\ Columbia University \\ New York, NY 10027}
\\ {\tt denef@math.columbia.edu}
\end{center}
\vfill

\begin{quote}

\begin{center}
{\bf Abstract}
\end{center}
\vskip 5mm

We investigate the correspondence between existence/stability of
BPS states in type II string theory compactified on a Calabi-Yau
manifold and BPS solutions of four dimensional N=2 supergravity.
Some paradoxes emerge, and we propose a resolution by considering
composite configurations. This in turn gives a smooth effective
field theory description of decay at marginal stability. We also
discuss the connection with 3-pronged strings, the Joyce
transition of special Lagrangian submanifolds, and $\Pi$-stability.

\end{quote}

\vfill


\end{titlepage}

\sect{Introduction}

An old and common idea in physics is that a particle makes its
presence manifest via excitation of fields. If one puts a lot of
particles together, one gets a macroscopic object, well described
by classical physics, and correspondingly one expects the field
excitations to be well described by a classical field theory. In
particular, it seems obvious that wherever we trust this field
theory as a good description of the low energy physics, a
well-behaved solution to the field equations corresponding to that
object should exist.

Type II string theory compactified on a Calabi-Yau manifold is
described at low energies by a four dimensional $\N=2$
supergravity theory coupled to massles vector- and
hypermultiplets. Quantum corrections to these theories are
relatively well under control, and yet they are remarkably rich in
content, with various intriguing connections to nontrivial physics
and mathematics.

When string perturbation theory can be trusted, massive charged
BPS particles in these theories can be described by D-branes
wrapping nontrivial supersymmetric cycles in the Calabi-Yau
manifold, or more generally by boundary states of the conformal
field theory describing the relevant string perturbaton theory.
When the low energy supergravity theory can be trusted, the same
objects can be described by solutions to the field equations of
motion. It turns out that not all charges support BPS states in
the string theory, and that not all charges have BPS solutions in
the supergravity theory. Thus, in suitable regimes, one naturally
expects some sort of correspondence between supergravity solutions
and D-branes.

Such a correspondence, while physically quite plausible, is in its
consequences highly nontrivial. For instance, it would give rise
to a number of powerful predictions about the existence of special
Lagrangian submanifolds in Calabi-Yau manifolds, and the existence
of boundary states in conformal field theories. However, as we
will show in this paper, closer examination of this supposed
correspondence reveals some intriguing puzzles. In particular, it
turns out that the traditional assumption of the particle as a
source localized in a single point of space leads to
inconsistencies. Fortunately, once again, string theory finds its
way out, and an interesting resolution to this paradox emerges.

The outline of this paper is as follows. In section 2, we briefly review the
relevant geometry underlying low energy type IIB string theory compactified on a
Calabi-Yau manifold. In section 3, the derivation of the attractor flow equations  
is revisited. We start from a duality invariant bosonic action, discuss  
an interpretation as a static string action, derive the spherically symmetric
attractor flow equations in two different forms, and comment on a subtlety
arising for vanishing cycles. In section 4, we analyze some properties of solutions,
with special
emphasis on conifold charges, leading to ``empty holes'', and a short discussion
of equal charge multicenter solutions. Then we tackle the existence issue: 
the attractor flow turns out to break down when the central charge has a
regular zero, and this leads to a natural conjecture on the existence of BPS
states \cite{M}. However, this natural conjecture
leads to some puzzling paradoxes. This is illustrated in section 5, using the
example of a certain BPS state at the Gepner point of the quintic, known to
exist, but nevertheless having a regular zero of the central charge. A second
puzzle is illustrated in the $SU(2)$ Seiberg-Witten model. In section 6, we propose 
a resolution to these puzzles; the key is to consider composite configurations. 
A thought experiment brings us rather naturally to the required configurations,
in a spherical shell approximation. A stability check is made using test
particle probes, and a representation as composite flows is given, making direct
contact with 3-pronged strings in a suitble rigid limit. A smooth effective
field theory picture for decay at marginal stability emerges, and Joyce's
stability conjecture for special Lagrangian submanifolds is recovered. We
briefly comment on $\Pi$-stability \cite{DFR}. Finally, in section 7, the analysis
of general stationary multicenter solutions is initiated. Some properties of
solutions can be inferred directly from the equations of motion. In particular,
the intrinsic angular momentum of the multicenter composites is computed. 
Section 8 summarizes our conlusions, and indicates some open problems.

\sect{Geometry of IIB/CY compactifications}

To establish notation and our setup, let us briefly review the low
energy geometry of type IIB string theory compactified on a
Calabi-Yau 3-fold. Some of the more technical elements of this
section are only needed for the derivation of some more technical
results further on.

We will follow the manifestly duality invariant formalism of
\cite{M}. Consider type IIB string theory compactified on a
Calabi-Yau manifold $X$. The four dimensional low energy theory is
$\N = 2$ supergravity coupled to $n_v = h^{1,2}$ massless abelian
vectormultiplets and $n_h = h^{1,1}+ 1$ massless hypermultiplets,
where the $h^{i,j}$ are the Hodge numbers of $X$. The
hypermultiplet fields will play no role in the following and are
set to zero.

The vectormultiplet scalars are given by the complex structure
moduli of $X$, and the lattice of electric and magnetic charges is
identified with $H^3(X,\IZ)$, the lattice of integral harmonic
$3$-forms on $X$. The ``total'' electromagnetic field strength
$\IF$ is (up to normalisation convention) equal to the type IIB
self-dual five-form field strength, and is assumed to have values
in $\Omega^2(M_4) \otimes H^3(X,\IZ)$, where $\Omega^2(M_4)$
denotes the space of 2-forms on the four dimensional spacetime
$M_4$. The usual components of the field strength are retrieved by
picking a symplectic basis ${\alpha^I,\beta_I}$ of $H^3(X,\IZ)$:
\begin{equation}
 \IF = F^I \otimes \beta_I - G_I \otimes \alpha^I. \label{FGcomp}
\end{equation}
The total field strength satisfies the self-duality constraint:
\begin{equation} \label{selfdual}
 \IF = *_{10} \IF,
\end{equation} where $*_{10}$ is the Hodge star
operator on the ten-dimensional space time, which factorises on
the $M_4 \times X$ compactification as $*_{10} = *_4 \otimes *_X$.
To prevent overly heavy notation, we will also denote the Hodge
dual in $X$ by a hat, so for any form $\Gamma$ on $X$:
\begin{equation}
 \widehat{\Gamma} \equiv *_X \Gamma
\end{equation}
Note that this operation is moduli-dependent. The constraint
(\ref{selfdual}) relates the $F$ and $G$ components in
(\ref{FGcomp}). The (source free) equation of motion and the
Bianchi identity of the electromagnetic field are combined in the
equation $\dop \IF = 0$, implying locally the existence of a
potential: $\IF = d \IA$.

The geometry of the vector multiplet moduli space, parametrized
with $n_v$ coordinates $z^a$, is special K\"ahler \cite{SG}. The
(positive definite) metric
\begin{equation}
  g_{a\bb} = \partial_a \bpartial_{\bb} \CK
\end{equation}
is derived from the K\"ahler potential
\begin{equation}
 \CK = - \ln ( \ii \int_X \Omega_0 \wedge \bOmega_0 ),
\end{equation}
where $\Omega_0$ is the holomorphic $3$-form on $X$, depending
holomorphically on the complex structure moduli. It is convenient
to introduce also the normalized 3-form
\begin{equation} \label{Omdef}
 \Omega = e^{\CK/2} \, \Omega_0.
\end{equation}
The ``central charge'' of $\Gamma \in H^3(X,\IZ)$ is given by
\begin{equation} \label{Zdef}
 Z(\Gamma) \equiv \int_X \Gamma \wedge \Omega \equiv \int_\Gamma
 \Omega,
\end{equation}
where we denoted, by slight abuse of notation, the cycle
Poincar\'e dual to $\Gamma$ by the same symbol $\Gamma$.

In the following we will frequently make use of the
(antisymmetric, topological, moduli independent) intersection
product:
\begin{equation} \label{intproddef}
\langle \Gamma_1,\Gamma_2 \rangle = \int_X \Gamma_1 \wedge
\Gamma_2 = \#(\Gamma_1 \cap \Gamma_2)
\end{equation}
With this notation, we have for a symplectic basis $\{
\alpha^I,\beta_I \}$ by definition $\langle \alpha^I,\beta_J
\rangle = \delta^I_J$. We will also often use the (symmetric,
positive definite, moduli dependent) Hodge product:
\begin{equation} \label{hodgeproddef}
\langle \Gamma_1,\widehat{\Gamma_2} \rangle = \langle\Gamma_1,*_X
\Gamma_2\rangle = \int_X \Gamma_1 \wedge *_X \Gamma_2.
\end{equation}
When the $\Gamma_i$ denote cohomology classes, their harmonic
representative will always be assumed in (\ref{hodgeproddef}).

Every harmonic $3$-form $\Gamma$ on $X$ can be decomposed
according to $H^3(X,\IC) = H^{3,0}(X) \oplus H^{2,1}(X) \oplus
H^{1,2}(X) \oplus H^{0,3}(X)$ as (for real $\Gamma$):
\begin{equation}
 \Gamma = i \bZ(\Gamma) \, \Omega \, - \, i g^{a\bb} \bar{D}_{\bb} Z(\Gamma) \,
 D_a \Omega \, + \, c.c. \, ,
\end{equation}
where we introduced the K\"ahler covariant derivative on $Z$ and
$\Omega$:
\begin{equation}
D_a \equiv \partial_a + \frac{1}{2} \partial_a\CK
\end{equation}
This decomposition is orthogonal with respect to the intersection
product (\ref{intproddef}), and diagonalizes the Hodge star
operator:
\begin{equation} \label{hodgediag}
 *_X \Omega = -i \, \Omega \quad \mbox{ and } \quad *_X D_a \Omega = i
\, D_a \Omega
\end{equation}

For further reference, we write down the following useful
identities:
\begin{eqnarray}
 \int_X \Omega \wedge \bar{\Omega} & = & -i \label{calculus1} \\
 \int_X D_a \Omega \wedge \bar{D}_{\bb} \bOmega &=& i \, g_{a \bb}  \label{calculus2} \\
 (d + i \, Q + i \, d\alpha) \,  (e^{-i \alpha} \Omega) &=& e^{-i \alpha} \, D_a \Omega \,
 dz^a \, , \label{calculus3}
\end{eqnarray}
where $\alpha$ is an arbitrary real function and $Q$ is the chiral
connection:
\begin{equation} \label{Qdef}
 Q = \Im (\partial_a {\cal K} dz^a) \, .
\end{equation}

As an example of an application, one can easily check the
following expressions for intersection and Hodge products:
\begin{eqnarray}
 \langle\Gamma_1,\Gamma_2\rangle &=& 2 \, \Im[- Z(\Gamma_1) \,
 \bar{Z}(\Gamma_2) \,+\, g^{a\bar{b}} \, D_a Z(\Gamma_1) \,
 \bar{D}_{\bar{b}} \bar{Z}(\Gamma_2)] \label{intproduct} \\
 \langle \Gamma_1, \widehat{\Gamma_2} \rangle &=& 2 \, \Re[Z(\Gamma_1) \,
 \bar{Z}(\Gamma_2) \,+\, g^{a\bar{b}} \, D_a Z(\Gamma_1) \,
 \bar{D}_{\bar{b}} \bar{Z}(\Gamma_2)] \label{hodgeproduct},
\end{eqnarray}

\sect{The attractor flow equations revisited} \label{revisit}

We now turn to the investigation of 4d supergravity BPS solutions
with charged sources corresponding to D3-branes wrapped around a
nontrivial supersymmetric (i.e. special Lagrangian) 3-cycle
\cite{BBS} $\Gamma$ of $X$. In the mirror IIA picture this
corresponds to BPS states with (mixed) 0-, 2-, 4- and 6-brane
charge.

Such $\N=2$ supergravity solutions and the remarkable attractor
mechanism emerging in this context were first studied, from
supersymmetry considerations, in \cite{FKS}. An approach based on
the bosonic action, which we will also follow here, was pioneered
in \cite{FGK}. Further explorations were made in \cite{attrsusy},
and various solutions analyzed in
\cite{attrsols,moresols,BLS,awg}. A rich connection with D-branes,
geometry and arithmetic was discovered in \cite{M}. Some recent
work on analogous phenomena in five dimensional theories includes
\cite{att5d}.

Part of this section is a review of well known results, though the
geometric, manifestly duality invariant setup we use may give a
clarifying alternative point of view on some of these. Also, the
strategy outlined here to obtain the BPS equations directly from
the bosonic action will enable us in section \ref{compflows} to do
the same for the general stationary case (possibly non-static,
with multiple centers having mutually nonlocal charges), adding
further insight to the solutions of \cite{BLS}. Furthermore, some
subtleties in the derivation of the flow equations will turn out
to be relevant for a proper treatment of the solution for conifold
charges later on, and finally, an interpretation of the reduced
action as that of an effective stretched string will allow us to
make contact with the $3-1-7$ brane description of BPS states in
$\N=2$ QFT.

So we believe it's worthwhile to revisit this derivation. However,
the reader only interested in the resulting equations can safely
skip the derivation and proceed with section \ref{solutions}.

\subsection{Duality invariant formalism} \label{diform}

The relevant bosonic part of the usual 4d low energy effective
${\cal N}=2$ supergravity action is, in 4d Planck units:
\begin{equation}
 S_{4D}=\frac{1}{16 \pi} \int_{M_4} d^4 x \sqrt{-G} R \,
  - \, 2 g_{a\bb} \, \dop z^a \wedge *\dop \bz^{\bb} \,\,
  - \frac{1}{4 \gamma^2} \int_{M_4} F^I \wedge G_I \label{S4D}
\end{equation}
where $\gamma$ is a convention dependent number, $F^I = d A^I$ and
the $G_I$ are obtained from the $F^I$ using the selfduality
constraint (\ref{selfdual}). On the other hand, the bosonic 4d
spacetime part of the low energy effective action of a probe
D3-brane wrapped around a supersymmetric 3-cycle in the homology
class $\Gamma$, in a given background, is \cite{BBS,D0,thesis}:
\begin{equation} \label{source}
 S_{\Gamma} = - \int |Z(\Gamma)| ds \,
 + \, \frac{\sqrt{\pi}}{\gamma} \int \langle\Gamma,\IA\rangle,
\end{equation}
with $Z(\Gamma)$ as in (\ref{Zdef}), $d \IA = \IF$, and $\langle
\cdot,\cdot \rangle$ denoting the intersection product
(\ref{intproddef}). The integral is over the effective particle
worldline.

Combining (\ref{S4D}) and (\ref{source}), assuming $\Gamma$ to be
electric with respect to the choice of symplectic basis (that is,
$\Gamma$ is a linear combination of the $\alpha^I$), we see that
an electromagnetic field produced by such a source with charge
$\Gamma$ satisfies, for any spatial surface $S$ surrounding the
source:
\begin{equation} \label{flux}
 \int_S \IF = \sqrt{4\pi} \gamma \, \Gamma
\end{equation}

Now while the action (\ref{S4D}) makes four dimensional general
covariance manifest, it is not invariant under electromagnetic
duality rotations (i.e. change of symplectic basis in
(\ref{FGcomp})). A straightforward, manifestly covariant action
exhibiting manifest duality invariance does not exist. In fact,
since the 4D theory descends directly from type IIB supergravity,
this problem is equivalent to the nonexistence of a
straightforward generally covariant action for the self-dual
four-form potential. However, a perfectly satisfactory, though not
manifestly covariant action for self-dual forms has been known for
quite a while \cite{HT}, and this action (dimensionally reduced)
will actually turn out to be very convenient for our purposes.

To write down this action for an arbitrary background metric, one
introduces the usual shift and lapse vectors \cite{MTW} $N_0$ and
$N^i$, putting the four dimensional metric in the form:
\begin{equation}
ds^2 = - N_0^2 dt^2 + G_{ij} (dx^i + N^i dt)^2.
\end{equation}
The shift vector determines a three dimensional 1-form $\sN =
G_{ij} N^j dx^i$. We will use boldface notation to refer to three
dimensional quantities throughout. Thus $\sd = dx^i \,
\partial_i$, the 3d Hodge dual (based on $G_{ij}$) is denoted by
$\s*$, and the spatial part of the total electromagnetic field
$\IF$ is
\begin{equation}
 \sF = \IF_{ij} \, dx^i \wedge dx^j = \sd \, \sA.
\end{equation}
The $H^3(X,\IR)$-valued 3-vector potential $\sA$ is considered to
be the fundamental variable (instead of the 4-vector potentials
$A^I$ in the formulation based on (\ref{S4D})). The action
obtained from \cite{HT} with our compactification assumptions is
then:
\begin{equation} \label{dualinvarS}
  S_{e.m.} = \frac{1}{4 \gamma^2} \int dt \int_{M_3}  \int_X \sF
  \wedge \partial_t \sA - (N_0 \, \sF \wedge \s* \widehat{\sF}
  + \sN \wedge \s* \sF \wedge \s* \sF)
\end{equation}
The integral over $X$ yields simply the intersection product
(\ref{intproddef}). Since the above expression does not refer to
any choice of symplectic basis, it is indeed manifestly duality
invariant. The equation of motion following from this action is
the self-duality condition (\ref{selfdual}), with $\IF=d\IA$,
where $\IA_0$ arises as an integration constant.

Of course, since $\IA_0$ does not exist off shell in this
formulation, we can no longer use the coupling of the
electromagnetic field to sources as in (\ref{source}). Instead,
its coupling to charges is implemented by imposing the constraint
(\ref{flux}), which only involves the spatial fields. Again, no
reference to a choice of basis is made. Note that the presence of
charges will induce Dirac string singularities in $\sA$, or
require the introduction of a nontrivial bundle structure.

The coupling of the source to gravity and the scalars remains
unchanged.

We will use (\ref{dualinvarS}) instead of the electromagnetic part
of (\ref{S4D}). In section \ref{compflows} the full form of this
action at nonzero $\sN$ will be used to derive the BPS equations
for the general stationary case, but it's instructive (and
sufficient for most of our purposes) to first consider some
simpler cases.

\subsection{Reduced action for static spherically symmetric
configurations}

In \cite{tod} it was argued that time independent BPS
configurations require a metric that can be expressed in the form
\begin{equation} \label{stationarymetric}
  ds^2 = - e^{2U} (dt + \omega_i dx^i)^2 + e^{-2 U} dx^i dx^i
\end{equation}
This is the metric ansatz we will use throughout this paper.
Usually we will also restrict to asymptotically flat spacetimes,
i.e. $U, \omega \to 0$ at spatial infinity. Let us first consider
static, spherically symmetric configurations. Then $\omega = 0$
and $U$ is a function of the radial coordinate $r=|\sx|$ only.
Similarly we take the moduli $z^a$ to be function of $r$ only, and
we can assume $\sF$ to be of the form
\begin{equation} \label{ssF}
 \sF = \frac{\gamma}{\sqrt{4\pi}} \sin \theta \, d\theta \wedge
 d\phi \, \otimes \, \Gamma
\end{equation}
where $\theta$ and $\phi$ are the usual angular coordinates and
$\Gamma \in H^3(X,\IZ)$ is the charge of the source. Then the
total electromagnetic field is, with $\tau \equiv 1/r$:
\begin{equation} \label{fullemfield}
  {\cal F} = \sF + *_4 \widehat{\sF} = \frac{\gamma}{\sqrt{4\pi}}
  ( \sin \theta \, d\theta \wedge
 d\phi \, \otimes \, \Gamma \, + \,
  e^{2U} d\tau \wedge dt \, \otimes \, \widehat{\Gamma})\, ,
\end{equation}
which trivially satisfies the required equations of motion and
duality constraints $d {\cal F} = 0$ and ${\cal F}=*_4
\widehat{\cal F}$.

In terms of the inverse radial coordinate $\tau = 1/r$, the total
action per unit time reduces, under these assumptions, and
dropping a total derivative proportional to $\ddot{U}$, simply to:
\begin{equation} \label{totenh1}
 S_{eff} = S/\Delta t = -\frac{1}{2} \int_0^\infty \dop \tau \; \{
 \dot{U}^2 + g_{a\bb} \, \dot{z}^a \dot{\bar{z}}^{\bb} +
 e^{2U} V(z)\} \, \, - (e^U |Z|)_{\tau=\infty}
\end{equation}
where the dot denotes derivation with respect to $\tau$ and (cf.
(\ref{hodgeproduct}))
\begin{eqnarray} \label{Vdef}
 V(z) &=& \frac{1}{2} \langle \Gamma,\widehat{\Gamma} \rangle \\
 &=& |Z|^2 + g^{a\bb} \, D_a Z \, \bar{D}_{\bb} \bZ
 = |Z|^2 + 4 g^{a\bb} \, \partial_a |Z| \, \bar{\partial}_{\bb}
 |Z|
\end{eqnarray}
with $Z=Z(\Gamma)$. The ``potential'' $e^{2U} V(z)$ is
proportional to the electromagnetic energy density. The boundary
term in (\ref{totenh1}) comes from (\ref{source}). In principle,
this reduced action has to be supplemented by the constraints
coming from variations of the metric (consistent with spherical
symmetry) other than the $U$ mode. In particular here this gives
the constraint $\dot{U}^2 + \| \dot{z} \|^2 - e^{2U} V(z) = 0$.
However, as we will see, this turns out to follow already from
(\ref{totenh1}) (with the given source coupling and allowing
arbitrary variations of the fields at $\tau = \infty$ in the
variational principle). So we will simply proceed with the
analysis of the action as it stands.

\begin{figure}
  \epsfig{file=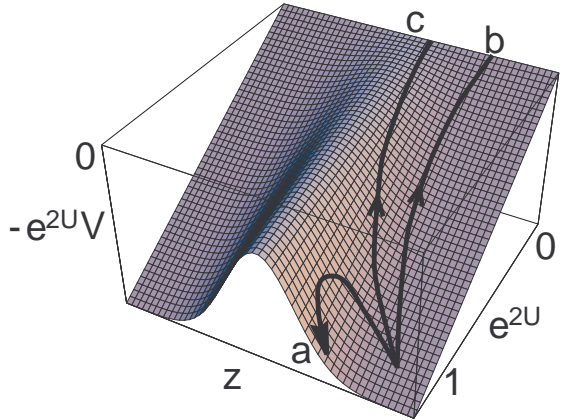,height=6cm,angle=0,trim= -120 0 0 0}
  \caption{Typical case sketch of the potential $-e^{2U} V$ in which the effective particle is
  moving, plotted as a function of $e^{2U}$ and the moduli $z$. Of
  the three plotted trajectories, only (c) satisfies the required
  asymptotic conditions yielding a BPS black hole.}
  \label{BHinvV}
\end{figure}

Note that (minus) this effective action per unit time can be
interpreted as describing a nonrelativistic particle moving in
$(U,z)$-space, subject to the potential $-e^{2U} V(z)$, with time
$\tau$ \cite{FGK} (fig.\ \ref{BHinvV}).
Only the initial ($\tau=1/r=0$) position of this effective
particle is fixed. The $\tau \to \infty$ asymptotic behavior is
given by requiring vanishing of the boundary terms at
$\tau=\infty$ when varying the fields. This yields for $\tau \to
\infty$
\begin{eqnarray}
 \dot{U} &\to& - e^U |Z| \label{as1} \\
 \dot{z}^a &\to& -2 e^U g^{a\bb} \, \bpartial_{\bb} |Z|. \label{as2}
\end{eqnarray}
Incidentally, these are precisely the attractor flow equations, as
we will see below. Note that this condition also implies the
vanishing of the (conserved) effective particle's energy, $E_{eff}
= \dot{U}^2 + \| \dot{z} \|^2 - e^{2U} V(z) = 0$, which is
precisely the ``additional'' constraint discussed earlier.

\subsection{Interpretation as a static string action}
\label{staticstring}

In fact, the reduced action (\ref{Vdef}) can also be interpreted
--- and this is perhaps more natural --- as a Polyakov
action for a static string in $(U,z)$ space, with (variable)
tension $T=e^U \sqrt{V(z)}$, in the background target space metric
\begin{equation}
 ds^2 = - dt^2 + dU^2 + g_{a\bb} dz^a d{\bar{z}}^{\bb},
\end{equation}
and worldsheet metric $\sim \mbox{diag}(-T,T^{-1})$ with respect
to the worldsheet time resp. space coordinates $t$ and $\tau$. The
vanishing of the effective particle's energy $E_{eff}$ is
equivalent to the Virasoro constraint, which can be used to
transform the action to the Nambu-Goto form
\begin{equation}
 S = - \int dt\, d\tau \, e^U \sqrt{V} \sqrt{\dot{U}^2+\| \dot{z} \|^2}.
\end{equation}
The asymptotic condition (\ref{as1})-(\ref{as2}), being equivalent
to the attractor flow equations, forces the endpoint of the string
at $\tau=\infty$ to be fixed at an attractor point (see below).
The other endpoint is fixed at the values of the moduli and $U$ at
spatial infinity in the supergravity picture. The equations of
motion determine the string to be a geodesic in $(U,z)$ space.

This interpretation, in a suitable rigid (i.e. gravity decoupling)
limit of $\N=2$ supergravity \cite{thesis,CYK3,awg} (leading to
Seiberg-Witten theory \cite{SW}\footnote{For an analysis of BPS
solutions of pure low energy effective $\N=2$ SU(2) Yang-Mills
theory, see \cite{CRU}. The solutions obtained there can be seen
to be the rigid limit of supergravity attractor flows \cite{awg}.}
and its generalizations), makes direct contact with the
description \cite{BPS37} of BPS states in $\CN=2$ quantum field
theories as stretched strings in a nontrivial background. For
example, repeating the above analysis for a charge $(n_e,n_m)$
state in the pure $SU(2)$ Seiberg-Witten effective theory (or
considering a suitable rigid limit of supergravity), one finds an
effective reduced action with similar structure, with one modulus
$u$, $V(u) = |n_e+n_m\tau|^2 / \Im \tau$ (where $\tau(u)$ is the
usual modular parameter of the SW Riemann surface), and evidently
$U \equiv 0$. For, say, a magnetic monopole, the attractor point
turns out to be its vanishing mass point $u=1$ (see also section
\ref{puzzle2}). Hence our effective string will be a geodesic
stretched in the Seiberg-Witten plane, between an arbitrary
modulus $u_{\tau=0}$ and $u=1$. Thus we arrive precisely at the
picture of \cite{BPS37}.

We will return to this point later on. In particular, the
phenomenon of three-pronged strings \cite{threeprong} appearing in
this context will turn out to be related to the resolution of some
intriguing paradoxes.

\subsection{BPS equations of motion} \label{bpseom}

The BPS equations of motion can be obtained from (\ref{totenh1})
by the usual Bogomol'nyi trick of completing squares. This can be
done in two ways, yielding two different forms of the BPS
equations. The first way is well known \cite{FGK}:
\begin{eqnarray}
 S_{eff} &=& - \frac{1}{2} \int_0^\infty \dop \tau \; \{ (\dot{U} \pm
 e^U |Z|)^2 + \| \, \dot{z}^a \pm 2 e^U g^{a\bb}
 \bpartial_{\bb}|Z|\,
 \|^2 \} \nonumber
 \\ &&\pm \, e^U |Z| \; \Big|^{\tau = \infty}_{\tau = 0} - (e^U
 |Z|)_{\tau=\infty}
 \label{kwadr_toten}
\end{eqnarray}
leading to the BPS equations
\begin{eqnarray}
 \dot{U} &=& - e^U |Z| \label{at1} \\
 \dot{z}^a &=& -2 e^U g^{a\bb} \, \bpartial_{\bb} |Z|. \label{at2}
\end{eqnarray}
This is the form of the equations found in \cite{FGK}. The other
sign possibility is excluded by the asymptotic condition
(\ref{as1})-(\ref{as2}), or alternatively, by requiring physical
acceptability: ``wrong sign'' solutions have runaway behavior,
severe curvature singularities at finite distance, negative ADM
mass, and are gravitationally repulsive.

The second way of squaring the action uses the Hodge product
(\ref{hodgeproddef}); if for a 3-form $C$ on $X$ we write
\begin{equation}
 \hn C \hn^2 = \langle C,\widehat{C} \rangle
\end{equation}
and we denote the position dependent phase of the central charge
as
\begin{equation} \label{alphadef}
 \alpha = \arg Z(\Gamma)
\end{equation}
then we have, using (\ref{calculus1})-(\ref{calculus3}):
\begin{eqnarray}
 S_{eff} &=& - \frac{1}{4} \int_0^\infty \dop \tau \;
 e^{2U} \, \hn \, 2 \, \Im[(\partial_\tau + i\, Q_\tau + i\, \dot{\alpha})
 (e^{-U} e^{-i \alpha} \Omega)] \, + \,
 \Gamma \, \hn^2 \nonumber \\
 &&- \, (e^U |Z|)_{\tau=0}  \label{square2}
\end{eqnarray}
where $Q_\tau = \Im(\partial_a {\cal K} \dot{z}^a)$, as in
(\ref{Qdef}). (We have left $U_{\tau=0}$ arbitrary here, though in
the asymptotically flat case this is zero of course.) Note that we
take the holomorphic 3-form $\Omega$ and the unnormalized
holomorphic $\Omega_0$ (cf.\ (\ref{Omdef})) to be only dependent
on the spacetime coordinates through the moduli $z^a(\tau)$. This
is in contrast to refs. \cite{attrsols}, where a (convenient)
explicit position dependence of normalisation and phase of
$\Omega_0$ was chosen.\footnote{One is free to make such a gauge
choice, at least locally, since phase and normalisation of
$\Omega_0$ do not enter the action (\ref{S4D}).} However, from a
Calabi-Yau geometrical point of view it is perhaps more natural to
pick a dependence only through the moduli. In numerical
computations, this has the further advantage that one can work
with a fixed expression for $\Omega$. Furthermore, in this way the
phase $\alpha$ appears naturally in the equations, and this phase
(which can be identified with the phase of the conserved
supersymmetry \cite{M}) will play a crucial role in the comparison
with geometrical results on special Lagrangian manifolds
\cite{joyce}.

The BPS equation following from (\ref{square2}) is again obtained
by putting the square to zero:
\begin{equation} \label{bps2}
 2 \, \Im[(\partial_\tau + i\, Q_\tau + i\, \dot{\alpha})
 (e^{-U} e^{-i \alpha} \Omega)] = - \Gamma.
\end{equation}
However, by taking the intersection product with $\Gamma$ on both
sides of the equation, and using (\ref{alphadef}), it follows that
$Q_\tau + \dot{\alpha} = 0$, hence the BPS equation becomes simply
\begin{equation} \label{bps3}
 2 \, \partial_\tau \, \Im (e^{-U} e^{-i \alpha} \Omega) = -
 \Gamma.
\end{equation}
Conversely, by taking the intersection product of the latter
equation with $\bar{\Omega}$, using (\ref{calculus3}) and
(\ref{calculus1}), and taking the imaginary part of the result,
one obtains again $Q_\tau + \dot{\alpha} = 0$, and hence
(\ref{bps2}). Now (\ref{bps3}) readily integrates to
\begin{equation} \label{integrated}
 2 \, \Im (e^{-U} e^{-i \alpha} \Omega) = - \Gamma \, \tau
 \, + \, 2 \, \Im (e^{-U} e^{-i \alpha} \Omega)_{\tau=0}.
\end{equation}
This is a powerful result, as it {\em solves}, in principle, the
BPS equations of motion of the system. To bring it in a perhaps
more familiar form, choose a symplectic basis $\{\alpha^I, \beta_I
\}$ of $H^3(X,\IZ)$, write $\Gamma = - q_I \alpha^I + p^I
\beta^I$, define the holomorphic periods $X^I = \langle \alpha^I,
\Omega_0 \rangle$, $F_I=\langle \beta_I, \Omega_0 \rangle$, and
take intersection products of this basis with the above equation.
This gives:
\begin{eqnarray}
 2 \, e^{-U+\CK/2} \, \Im (e^{-i \alpha} X^I) = \frac{p^I}{r} + c^I \\
 2 \, e^{-U+\CK/2} \, \Im (e^{-i \alpha} F_I) = \frac{q_I}{r} + d_I \, ,
\end{eqnarray}
where we re-introduced $r=1/\tau$, and $c^I, d_I$ are constants.
If, as in \cite{attrsusy,attrsols}, we would pick an
$\Omega_0$-gauge where $\CK \equiv 2 U$ and $\alpha \equiv 0$, one
retrieves the expressions appearing in those references. Note also
that the flow equations (\ref{at1})-(\ref{at2}) are nothing but
the projections of (\ref{bps2}) on $e^{-i \alpha} \Omega$ resp.
$e^{-i \alpha} D_a \Omega$.

Of course, finding the explicit flows in moduli space from
(\ref{integrated}) requires inversion of the periods to the
moduli, which in general is not feasible analytically. However, in
large complex structure approximations \cite{attrsols} or
numerically for e.g. the quintic, this turns out to be possible.

One final remark is in order. The solution (\ref{integrated}) was
derived from the action (\ref{square2}) under the implicit
assumption that the quantity between brackets is not proportional
to (the Poincar\'e dual of) a vanishing cycle $\nu$ (that is, a
cycle for which the Hodge square $\langle \nu, \widehat{\nu}
\rangle = 0$, like the conifold cycle at a conifold point of
moduli space). If that is the case, the Hodge square appearing in
(\ref{square2}) is automatically zero, no matter what the
expression inside the $\hn \cdot \hn$ is (as long as it is
finitely proportional to a vanishing cycle). Actually, a we will
see, such situations {\em do} occur naturally, and the previous
remark should eliminate possible confusion there.

\sect{Attractors and existence of BPS states} \label{solutions}

\subsection{Properties of some solutions}

\begin{figure}
  \epsfig{file=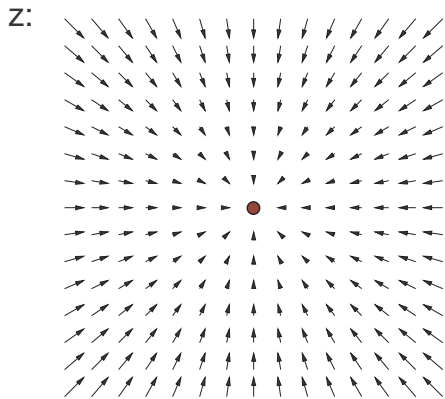,height=6cm,angle=0,trim= -100 200 0 0}
  \caption{Typical flow gradient field in moduli space (represented here by the
  $z$-plane) close to a generic attractor point with nonzero minimal $|Z|$. The gradient
  vectors vanish at the attractor point.}
  \label{BHflowvec}
\end{figure}

In what follows we will assume asymptotic flatness, i.e.\
$U_{\tau=0} = 0$. Then solutions to (\ref{at1})-(\ref{at2})
saturate the BPS bound
\begin{equation}
  M = |Z_{\tau=0}|. \label{BPS}
\end{equation}
All mass is located in the fields: the ``bare mass'' contribution
$(e^U |Z|)_{\tau=\infty}$ is zero. Indeed, (\ref{at1}) and
(\ref{at2}) imply that both $e^U$ and $|Z|$ are monotonously
decreasing functions satisfying the estimate $e^U |Z| \leq |Z|/(1
+ |Z_\infty| \tau)$, hence $e^U |Z| \to 0$ when $\tau \to \infty$.
More precisely, equation (\ref{at2}) implies
\begin{equation}
 \partial_\tau |Z| = -4 e^{U} g^{a\bb}
 \; \partial_a |Z| \, \bpartial_{\bb} |Z| \leq 0 \, , \label{absZ}
\end{equation}
so the flows in moduli space described by (\ref{at1}) and
(\ref{at2}) converge to minima of the central charge modulus
$|Z(\Gamma)|$ (fig.\ \ref{BHflowvec}).
Therefore the moduli values at the horizon are generically
invariant under continuous deformations of the moduli at spatial
infinity, and hence only dependent on the charge $\Gamma$, a
phenomenon referred to as the attractor mechanism. The attractor
values of the moduli correspond to Calabi-Yau manifolds with very
remarkable arithmetic properties, as explored in detail in
\cite{M}.

\subsubsection{Black holes}

The above estimate also implies that when $Z_\infty \neq 0$, the
solution is a black hole with horizon at $\tau=\infty$. From the
form of the metric (\ref{stationarymetric}) and direct analysis of
equation (\ref{at1}) in the limit $\tau \to \infty$, the near
horizon geometry can be seen to be $AdS_2 \times S^2$:
\begin{equation}\label{NHmetric}
 ds^2_{NH} = - \frac{r^2}{{|Z_\infty|}^2} dt^2
 + \frac{{|Z_\infty|}^2}{r^2} d {\bf x}^2.
\end{equation}
The corresponding macroscopic entropy is $A/4=\pi |Z_\infty|^2$.
These black holes have been studied extensively in the literature
\cite{FKS,FGK,attrsusy,attrsols,BLS,moresols}.

\subsubsection{Empty holes} \label{emptyholes}

When the D3-brane wraps a conifold cycle, i.e. a cycle vanishing
at a conifold point, the minimal value of the central charge
modulus is zero, and the above discussion of the generic case does
no longer apply. However, since conifold cycles are known to exist
(close to a conifold point) as special Lagrangian submanifolds,
and therefore as physical BPS particles, it is natural to ask what
the corresponding supergravity solution looks like.

Again, the flow in moduli space will converge to a point where
$|Z|$ is minimal, which in this case is a point on the conifold
locus, where $Z$ vanishes. At the conifold locus, the Calabi-Yau
degenerates and we get an additional massless hypermultiplet in
the low energy theory, so there we cannot necessarily trust our
supergravity approximation. However, the results obtained are
physically pleasing and interesting, so we will ignore this
potential problem and proceed.

For simplicity, following \cite{M}, we will consider only one
modulus, $z$, which we define to be the holomorphic ($\Omega_0$)
period of the vanishing cycle. Then the K\"ahler potential and
metric close to the conifold point ($z \to 0$) can be taken to be:
\begin{eqnarray}
 e^{-\CK} &\approx& k_1^2 + \frac{1}{2 \pi} |z|^2 \ln |z|^2 + k_2 \Re z
 \label{cyberradio} \\
 g_{z\bar{z}} &\approx& \frac{1}{2 \pi k_1^2} \ln |z|^{-2}, \label{CM}
\end{eqnarray}
where $k_1$ and $k_2$ are positive constants. This geometry can be
observed for instance close to the conifold point of the quintic,
or (in a rigid limit) close to the massless monopole or dyon
singularities in Seiberg-Witten theory.

The central charge of $N$ times the vanishing cycle close to $z=0$
is $Z = \frac{N}{k_1} z$ and the attractor flow equations in this
limit are
\begin{eqnarray}
  \dot{U} &=& - \frac{N}{k_1} e^U |z| \\
  \dot{z} &=& - 2 \pi k_1 N e^U \frac{z}{|z|} \frac{1}{\ln
  |z|^{-2}} \, , \label{coneq2}
\end{eqnarray}
with solution (approximately for $z \to 0$) given by:
\begin{eqnarray}
  \arg z &=& \mbox{const.} \label{emsol1} \\
  |z| \ln |z|^{-1} &=& \left\{
  \begin{array}{lcl}
  \pi k_1 N \, e^{U_*} \, (\tau_*-\tau) & \quad \mbox{for }& \tau < \tau_* \\
  0 & \quad \mbox{for } & \tau \geq \tau_*
  \end{array}
   \right. \label{emsol2} \\
  U &=& \frac{1}{4 \pi k_1^2} |z|^2 \ln |z|^{-2} + U_* \, , \label{emsol3}
\end{eqnarray}
where $\tau_*$ and $U_*$ are constants depending on initial
conditions.

\begin{figure}
  \epsfig{file=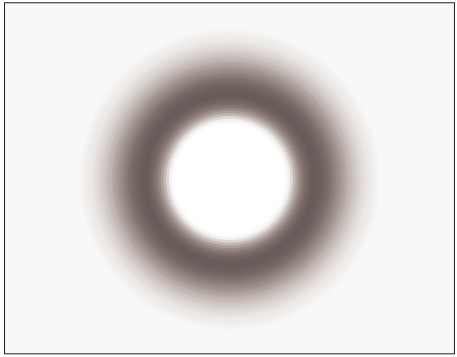,height=5cm,angle=0,trim= -250 0 0 0}
  \caption{Energy density sketch of an ``empty hole''.}
  \label{empty}
\end{figure}

So here the attractor point $z=0$ is reached at {\em finite
nonzero} coordinate distance $r_*=1/\tau_*$ from the origin. In
the core region $r<r_*$, the fields $z$ and $U$ are constant and
the geometry is flat. There is no horizon and the core contains no
energy ($\dot{U}=\dot{z}=0$ and $V(z)=0$), hence the name ``empty
hole'' (fig. \ref{empty}).
The solution $z(\tau)$ is once and $U(\tau)$ twice continuously
differentiable at $\tau=\tau_*$, so the curvature stays finite
(and can be made arbitrary small by taking $N$ sufficiently
large). However, higher derivatives diverge at $\tau=\tau_*$, so
strictly speaking the (two derivative) supergravity approximation
breaks down here.\footnote{This can perhaps be cured by including
higher derivative terms or the new massless hypermultiplet in the
effective action, smoothing out the solution, but presumably not
changing it too much.} Nevertheless, we believe this is a
physically sensible solution. For instance, one could use it (in
the straightforward multicenter extension given section
\ref{multicentercase}) to compute the dynamics of a large number
of slowly moving empty holes, in moduli space approximation, with
sensible results. Note that due to the fact that these solutions
are {\em not} black holes, one will obtain a moduli space geometry
for nearly coincident centers which is completely different from
the black hole one discussed in \cite{modulispace,modulispace2}.
In particular, there will presumably be no coalescence, in
agreement with the physical expectation that no BPS {\em bound}
states should exist for branes wrapping a conifold cycle \cite{S}.

We could also have used (\ref{integrated}) to construct this
solution (in particular cases this would in fact be a more
powerful method to extract exact results, also away from the near
conifold limit). However, naive application of this formula would
lead to a field configuaration that is not constant inside the
core: at $\tau=\tau_*$, the central charge phase $\alpha$ jumps
discontinuously from $\alpha_*$ to $\alpha_* + \pi$, and for $\tau
> \tau_*$, one gets the ``solution'' corresponding to the flow
equations (\ref{at1})-(\ref{at2}) with the opposite (=wrong) sign.
As discussed in section \ref{bpseom}, this is not an acceptable
solution; it is not BPS, and physically ill-behaved.

The way out of this paradox is the remark given at the end of
section \ref{bpseom}: equation (\ref{integrated}) needs only to be
satisfied down to the radius where the conifold attractor point is
reached. If we keep the fields constant below this radius, the BPS
condition is automatically satisfied. This eliminates some
confusion arising in the literature in this context.

Note that even though the solution (\ref{emsol1})-(\ref{emsol3})
was derived in the near conifold limit, the conclusion that the
attractor point is reached at finite $\tau$ is also true for
moduli at infinity farther away from the conifold point, since the
region where the approximation becomes valid will in any case be
reached after finite $\tau$. Furthermore solutions at different
$N$ are related by simple scaling; the core radius is proportional
to $N$. So the solution will never be a black hole, no matter
where we start in the moduli space, and no matter how many
particles we put on top of each other. The attractor mechanism
causes the mass to stay outside the Schwarzschild radius,
protecting the configuration from gravitational collapse.

If the modulus $z_0$ at spatial infinity $\tau=0$ is sufficiently
small, the core radius is given by $r_*=\frac{\pi k_1 N }{|z_0|
\ln |z_0|^{-1}}$. In the zero mass limit $z_0 \to 0$, the core
radius goes to infinity, leaving a completely flat space. If on
the other hand one boosts up the particle while sending $z_0 \to
0$, in such way that the total energy remains constant, one
obtains \cite{awg} in the limit $z_0 \to 0$ the Aichelburg-Sexl
shockwave metric \cite{aisexl} for a massless particle moving at
the speed of light. Again, this is physically sensible.

Finally note that we have derived the empty hole solution assuming
all charge to be located at ${\bf x}=0$. However, exactly the same
solution for $U$ and the moduli would have been obtained for any
spherically symmetric charge distribution inside the core region.
In particular the energy density and space curvature would have
been the same. In that sense the charge is actually {\em
delocalized}. It could for example be a spherical shell of
radius $r_*$ (this is perhaps the most natural location of the
charges, as the ``emptyness'' of the core then becomes quite
intuitive).

All this is of course very reminiscent of the enhan\c{c}on
mechanism of \cite{enhancon}. One could say that the massless
conifold particle is the ``enhan\c{c}on'' curing the repulson
singularity one would obtain for example by applying naively
formula (\ref{integrated}). The main difference is that there is
no enhanced gauge symmetry in the core region, but rather an
additional massless charged hypermultiplet.

It would be interesting to find out whether empty holes, like
their black hole cousins \cite{adsfrag}, also have a Maldacena
dual \cite{AdSCFT} QFT description.

\subsubsection{No holes} \label{noholes}

\begin{figure}
  \epsfig{file=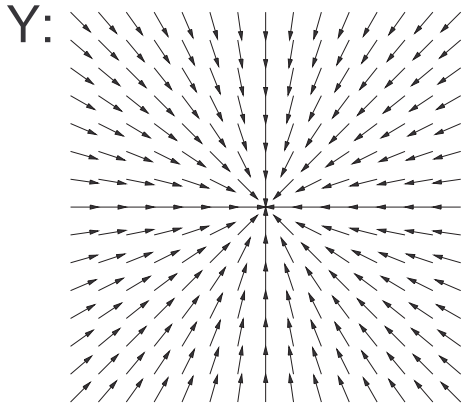,height=6cm,angle=0,trim= -100 200 0 0}
  \caption{Flow gradient field in the $Y=\int_{\Gamma} \Omega_0$-plane
  close to a regular point $Y=0$ in moduli space where the central charge vanishes.
  The gradient vectors do not vanish at the attractor point, leading to a breakdown
  of the flow.}
  \label{regflowvec}
\end{figure}

As observed in \cite{M}, the flow equations
(\ref{at1})-(\ref{at2}) do not always have a solution: if the
attractor point of the flow happens to be a simple zero of $Z$, at
a regular point of moduli space, the flow will reach $Z=0$ at
finite $\tau=\tau_*$ and {\em cannot} be continued in a BPS way to
the interior region $\tau>\tau_*$ (see also fig.\
\ref{regflowvec}).
The basic difference with the previous case is the absence of the
``damping'' factor $1/\ln |z|^{-2}$ in the inverse metric on the
right hand side of (\ref{at2}) (or (\ref{coneq2})), so that the
constant field configuration at $Z=0$ is no longer a solution. On
the other hand, by taking the charge sufficiently large, the
curvature can be made again arbitrary small, so the absence of a
supergravity solution should be quite meaningful.

Physically, one indeed doesn't expect a BPS state with charge
$\Gamma$ to exist in a vacuum near a regular point where
$Z(\Gamma)=0$: such a particle would be massless at $Z=0$, which
(by integrating it out) should lead to a singularity in moduli
space \cite{S}, in contradiction with the supposed regularity of
the point under consideration.

\subsection{Equal charge multicenter solutions}

\label{multicentercase}

\begin{figure}
  \epsfig{file=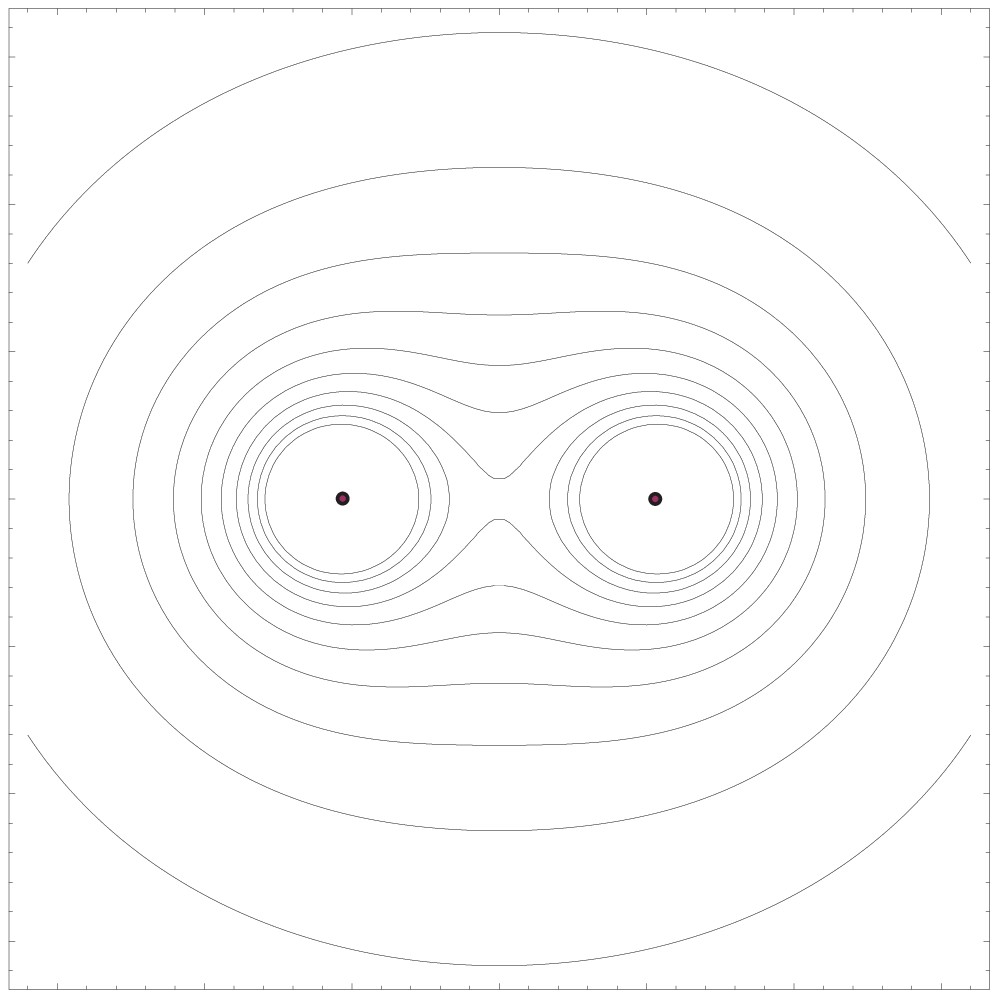,height=5cm,angle=0,trim=-300 0 0 0}
  \caption{Some surfaces of equal $\tau$ in the 2-center case.}
  \label{twocenter}
\end{figure}

The single center configuration discussed above is readily
extended to the multicenter case with equal (or parallel) charges
in the centers. (Multicenter solutions with non-parallel charges
are considerably more involved, and will be discussed in section
\ref{compflows}.) One simply replaces $\tau = 1/|{\bf x}|$ by
\begin{equation}
 \tau \equiv \frac{1}{N} \sum_{i=1}^N \frac{1}{| {\bf x} - {\bf x}_i |} \,
 ,
\end{equation}
where the ${\bf x}_i$ denote the positions of the particles, each
with charge $\Gamma$ (fig. \ref{twocenter}).
Since the complete setup is formally the same as for the
spherically symmetric case, so are the attractor flow equations.
Therefore, everything said about the spherically symmetric case
applies to the multicenter case as well.

For nearly coincident centers, the black hole near horizon
geometry now becomes ``fragmented'' $AdS_2 \times S^2$, as
discussed in \cite{adsfrag}.

Though a proof for the general case is still lacking, it is
expected \cite{GP} that the moduli space geometry for the dynamics
of slowly moving centers can be derived \cite{modulispace2} from
the potential
\begin{equation}
 L = \int d^3{\bf x} \, e^{-4U} \, .
\end{equation}

\subsection{Existence of BPS states} \label{existence}

The issue of existence of BPS states with given charge in theories
with ${\cal N}=2$ supersymmetry in four dimensions is nontrivial
and profound. The simplest example of such a theory is probably
$SU(2)$ $\N=2$ Yang-Mills theory. The low energy dynamics of this
theory was exactly solved in \cite{SW}, where it was found that
the BPS spectrum at weak coupling consists of the gauge boson and
a tower of dyons of arbitrary integer electric charge and one unit
of magnetic charge, while at strong coupling it consists solely of
the magnetic monopole and the ``elementary'' dyon with one unit of
electric and one unit of magnetic charge. Here ``strong coupling''
has the precise meaning of being inside a certain curve in the one
dimensional moduli space, called the curve of marginal stability.
At this curve, the various BPS particles have parallel central
charges, so that they become only marginally stable against decay
into constituents.

Similar phenomena are expected for type II string theory
compactified on a Calabi-Yau 3-fold. Here the subject is
intimately related to the existence of D-branes wrapped on
supersymmetric cycles, since these are the objects that represent
the BPS states. For instance in type IIB theory, at least in the
large complex structure limit, existence of a BPS state of charge
$\Gamma \in H^3(X,\IZ)$ is equivalent with existence of a special
Lagrangian submanifold in the homology class (Poincar\'e dual to)
$\Gamma$ \cite{BBS}. In type IIA theory, in the large volume
limit, it is equivalent with existence of holomorphic submanifolds
endowed with certain holomorphic bundles (or more precisely
sheaves). At certain special points in moduli space, existence can
be proven using the boundary state formalism. Recently, the
problem has been studied intensively from various points of view:
special Lagrangian submanifolds \cite{BBS,hitchin,joyce,KM,GJ},
holomorphic geometry and boundary states
\cite{RS,BDLR,D,DR,Sch,BS,DFR,DFR2} and low energy effective
supergravity \cite{M,D}.

We will study this problem from the latter point of view, namely
the low energy supergravity theory. The idea \cite{M} is as
follows. If a certain charge supports a BPS state, one certainly
would expect a corresponding 4d supergravity solution to exist, at
least for sufficiently large charge, such that the supergravity
approximation can be trusted. The converse statement is perhaps
less evident, but with the knowledge that some charges indeed do
not have BPS supergravity solutions (see section \ref{noholes}),
it is quite tempting to conjecture an exact correspondence, at
least for sufficiently large charges. Clearly, considering the
degree of difficulty of the problem in other approaches, such a
correspondence would be very powerful.

The above considerations were used in \cite{M} to arrive at the
following proposal for an existence criterion for BPS states with
given charge. Choose moduli ${z^a}_0$ at spatial infinity and a
charge $\Gamma$, and denote the minimal value of $|Z(\Gamma)|$
where the solution of (\ref{at1})-(\ref{at2}) flows to by
$|Z|_{min}$. There are three distinguished cases.
\begin{itemize}
 \item {\em Type a}: $|Z|_{min} \neq 0$. The attractor flow exists and
 yields a regular BPS black hole solution. In this case one
 expects to have a BPS state in the theory with the given charge.
 Note that if the existence of a BPS state in a certain vacuum
 ${z^a}_0$ is thus established, it will also exist
 in any other vacuum that lies ``upstream'' the $\Gamma$ attractor
 flow passing through the point ${z^a}_0$, where ``upstream''
 means in the opposite direction of the flow given by
 (\ref{at1})-(\ref{at2}). Since $|Z|$ has no maxima in moduli
 space \cite{FGK}, the upstream flows will tend to regions of
 moduli space at infinite distance, like the large complex
 structure limit. This also explains to a certain extent why BPS states are
 more likely to exist at large complex structure than in
 the bulk of modulispace.
 \item {\em Type b}: the flow tends to a singularity or a boundary of
 moduli space, where $|Z|$ might or might not vanish. More
 information is needed to decide whether the BPS state exists or
 not.
 \item {\em Type c}: $|Z|_{min}=0$, and this minimum is reached
 at a regular point in moduli space. As discussed in section
 \ref{noholes}, the flow breaks down and the charge is expected
 not to support a BPS state.
\end{itemize}

Though this proposed criterion works nicely for e.g.\ $T^6$
\cite{M}, it can fail in more general cases, as we will argue in
the next section. More precisely, it turns out that some type c
cases {\em do} correspond to BPS states present in the theory.


\sect{Puzzles and paradoxes}

\subsection{Puzzle 1: Solution suicide; states at the Gepner point of the quintic.}
\label{puzzle1}

We start by considering the example of the quintic Calabi-Yau,
first analyzed in great detail in \cite{cand}. In particular, we
will study BPS states in type IIB theory on the mirror quintic $W$
(or equivalently in type IIA on the quintic $M$ itself). This
manifold can be obtained \cite{GrP} as a $\IZ_5^3$ quotient of the
manifold in $\IP^4$ given by the equation
\begin{equation}
 W: x_1^5 + x_2^5 + x_3^5 + x_4^5 + x_5^5 - 5 \, \psi \, x_1 x_2 x_3 x_4
 x_5 = 0 \,.
\end{equation}
The transformation $\psi \to \omega \psi$ with $\omega^5=1$ can be
undone by a coordinate transformation $x_1 \to \omega^{-1} x_1$,
and thus the complex structure moduli space of $W$ can be
parametrized by $\psi^5$. The moduli space has three
singularities: the Gepner point $\psi^5=0$, which is a $\IZ_5$
orbifold singularity, the conifold point $\psi^5=1$, where a
3-cycle vanishes, and the large complex structure limit
$\psi^5=\infty$, mirror to the large volume limit of the quintic.

In \cite{BDLR}, building on \cite{RS},
the D-brane spectrum of this theory was studied,
mainly from the conformal field theory perspective. In particular
the existence of a number of BPS states was established at the
Gepner point $\psi=0$. These states were labeled as $|0 0 0 0
0\rangle_B, |1 0 0 0 0 \rangle_B, \ldots, |1 1 1 1 1 \rangle_B$.
The state $|0 0 0 0 0\rangle_B$ corresponds to a D3-brane wrapped
around the conifold cycle on the type $IIB$ side, and to a
D6-brane on the type $IIA$ side.\footnote{The identification of
the type IIA D-brane charges depends on the chosen analytic
continuation to large complex structure, so it has some intrinsic
arbitrariness (see \cite{BDLR} for some discussion of this
point).} It becomes massless at the conifold point $\psi=1$. The
state $|1 0 0 0 0\rangle_B$ has two units of D6-brane charge and
five units of D2-brane charge in the type IIA theory, and the
state $|1 1 0 0 0\rangle_B$ has one unit of D6- and five units of
D2-brane charge. The expected dimension of the deformation moduli
space of these three states is respectively 0, 4 and 11.

According to the existence criterion of section \ref{existence},
we should find ``good'' attractor flows with $\psi=0$ at spatial
infinity for all these states; they should not be of type c. To
address this question, one needs the exact moduli space geometry
and central charges ($\Omega$-periods) at arbitrary points in
moduli space for the charges under consideration. From the results
of \cite{cand,BDLR,GLaz}, all this is indeed available, in terms
of certain Meijer functions \cite{GLaz} of the modulus $\psi^5$.
It is still hard then to tackle this problem analytically, but
numerically using for instance Mathematica, it becomes quite
tractable.
\begin{figure}
  \epsfig{file=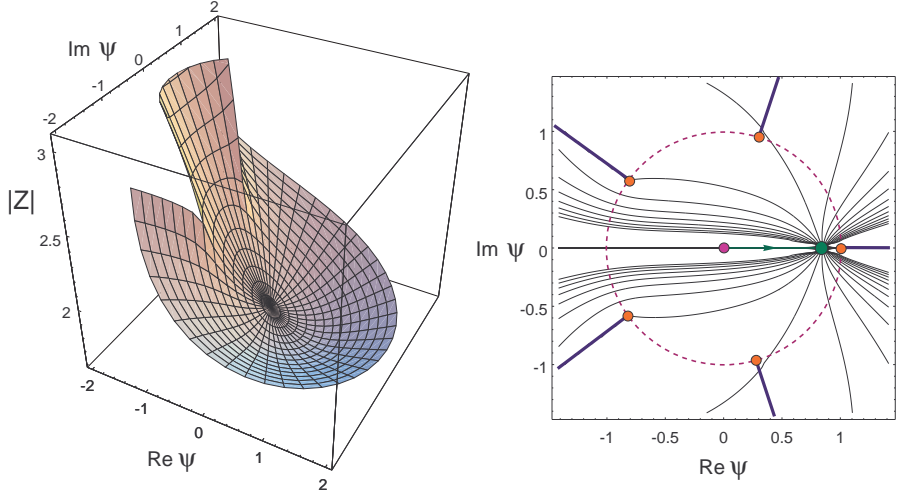,height=7cm,angle=0,trim=-30 0 0 0}
  \caption{{\em Left}: modulus of the central charge as a function of $\psi$
  for the state $|11000\rangle_B$. The variable $\psi$ parametrizes
  a five-fold cover of the moduli space around the Gepner point $\psi=0$.
  The discontinuities in the graph are due to
  monodromies around the conifold points $\psi^5=1$. There is a regular
  nonzero minimum $|Z|_{min} \approx 1.61$ at $\psi \approx 0.85$.
  {\em Right}: corresponding flows in the $\psi$-plane. The five fat blue lines
  are the cuts for the periods, starting at the conifold points.
  The green line with arrow from $\psi=0$ to the attractor point
  is the flow for the Gepner vacuum.}
  \label{11000}
\end{figure}

\begin{figure}
  \epsfig{file=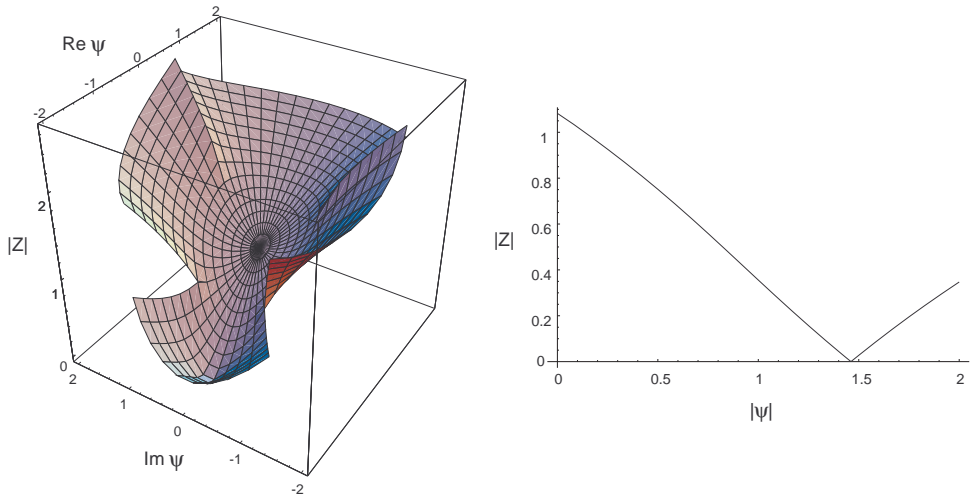,height=7cm,angle=0,trim=-30 0 0 0}
  \caption{{\em Left:} modulus of the central charge as a function of $\psi$
  for the state $|10000\rangle_B$ (notice the change in viewpoint with respect to
  fig.\ \ref{11000}: the $\psi$-plane is rotated $90$ degrees). There is a regular
  zero at $\psi \approx -1.46$. {\em Right:} $|Z|$ as a function of $|\psi|$ on the
  negative real axis.}
  \label{10000}
\end{figure}

As an example, we show in fig.\ \ref{11000} the modulus of the
central charge $Z$ as a function of $\psi$ for the state
$|11000\rangle_B$. In this case we find indeed a nice regular BPS
black hole solution, with $|Z|_{min} \approx 1.61$ at the
attractor point $\psi \approx 0.85$ (to make the supergravity
approximation valid, we should actually put a large number $N$ of
these charges on top of each other, but this simply rescales the
solution). The same is true for $|11100\rangle_B$ ($|Z|_{min}
\approx 2.78$ at $\psi \approx -0.51$), for $|11110\rangle_B$
($|Z|_{min} \approx 4.58$ at $\psi \approx -0.15$), and for
$|11111\rangle_B$ ($|Z|_{min} \approx 7.43$ at $\psi \approx
-0.07$). For $|00000\rangle_B$ we find an empty hole solution with
attractor point $\psi=1$.\footnote{Incidentally, in all cases we find
$|Z(\psi)|$ to be symmetric under $\psi \to \bar{\psi}$, illustrating the
rather special character of the boundary states constructed in \cite{RS,BDLR}.}

However, as also noticed in \cite{D}, for the state
$|10000\rangle_B$, we are in trouble. As shown in fig.\
\ref{10000}, the attractor point $\psi \approx -1.46$ is a regular
zero of $|Z|$, so we have a type c situation: the supergravity
solution does not exist!

Note that, though in conflict with the criterion of section
\ref{existence}, this result is {\em not} in contradiction with
the physical expectation that a charge with $Z=0$ at a regular
point cannot support BPS states {\em in a neighborhood} of that
point: the zero $\psi=\psi_* \approx -1.46$ and the Gepner point
$\psi=0$, where the existence of the state is established, can be
separated by a line of marginal stability where the state decays
into lighter BPS constituents. So it is perfectly possible to have
a BPS state with the given charge at $\psi=0$ and no such state
close to $\psi=\psi_*$.

\begin{figure}
  \epsfig{file=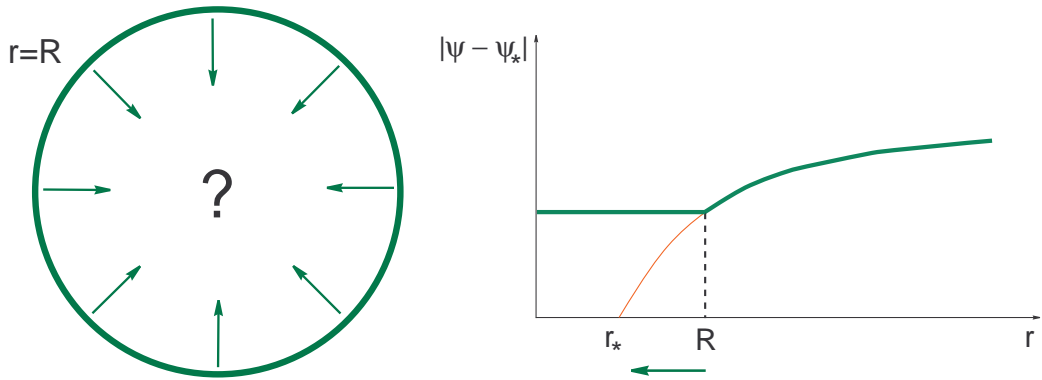,height=5.4cm,angle=0,trim=0 0 0 0}
  \caption{{\em Left:} shell of BPS particles with zero central charge at
  the regular point $\psi=\psi_*$, slowly
  moving inward. {\em Right:} sketch of $|\psi-\psi_*|$ as a
  function of the distance $r$ from the origin. The thick green line is
  the actual value when the radius of the sphere equals $R$. The
  thin red line shows how $|\psi-\psi_*|_{r=R}$ would progress when the
  shell moves further inward.}
  \label{collapse}
\end{figure}

Still, we clearly do have a physical problem here. This can be
seen most clearly by considering the following thought experiment
(fig.\ \ref{collapse}). Imagine a very large number of particles
with the given charge on a huge sphere of radius $R$, in a vacuum
with $\psi=0$. For $R \to \infty$ we expect to be allowed to
neglect the backreaction of the particles on the bulk fields, and
the description of these particles as CFT boundary states in the
fixed background should be valid. Since we know from \cite{RS,BDLR}
that the CFT boundary states corresponding to these particles
exist and are BPS in the given vacuum, such a configuration should
indeed exist and be BPS. Now give each of the particles the same
very small inward velocity, and let us approximate the particle
cloud as a uniformly charged spherical shell of adiabatically
decreasing radius $R$. When the sphere becomes smaller, the
collective backreaction becomes more important: outside the
sphere, the fields will be given by the attractor flows
(\ref{at1})-(\ref{at2}); inside the sphere, the fields are
constant. Note that this configuration is indeed BPS: the energy
stored in the bulk fields outside the shell is
$E_{out}=|Z|_{r=\infty} - (e^U |Z|)_{r=R}$, the energy of the
shell itself is $E_{shell}=(e^U |Z|)_{r=R}$, and the energy stored
in the fields inside the shell is zero, adding up to a total
energy $E_{tot} = |Z|_{r=\infty}$.

But if this motion goes on and nothing happens, we run into
disaster: when $R$ becomes smaller than the nonzero radius $r_*$
where the attractor flow breaks down, we no longer have a sensible
solution! Moreover, by the physical argument given earlier, we
actually expect that the particle cloud doesn't even exist anymore
at this point...

We propose a way out in section \ref{resolutions}.


\subsection{Puzzle 2: Monodromy murder; dyons in Seiberg-Witten theory}
\label{puzzle2}

For our second (but closely related) puzzle, we consider the
monopole in the Seiberg-Witten low energy effective theory for
$SU(2)$ ${\cal N=2}$ Yang-Mills \cite{SW}. This theory can be
obtained from the ${\cal N}=2$ supergravity theory describing the
low energy physics of type II string theory compactified on a
suitable Calabi-Yau manifold, in a certain rigid (=gravity
decoupling) limit \cite{geomeng,CYK3,awg,thesis}. The BPS
solutions of this effective abelian theory (see \cite{CRU} for a
discussion taking into account nonabelian corrections) can
correspondingly be obtained as rigid limits of supergravity
attractor flows \cite{awg}. Because gravity decouples, $U$ is zero
everywhere. The attractor flow equation for the modulus $u(\tau)$
is\footnote{The factor $\sqrt{2}$ is due to the conventions used
in \cite{SW}. We take $\Lambda \equiv 1$.}
\begin{equation} \label{atSW}
  \dot{u} = - \sqrt{2} \, g^{u\bar{u}} \, \partial_{\bar{u}} |Z|
  \, ,
\end{equation}
where $g^{u\bar{u}}$ is the inverse Seiberg-Witten metric and $Z$
is the central charge; for electric charge $n_e$ and magnetic
charge $n_m$, this is $Z=n_e a(u) + n_m a_D(u)$, where $a$ and
$a_D$ are given by certain hypergeometric functions \cite{SW}.

\begin{figure}
  \epsfig{file=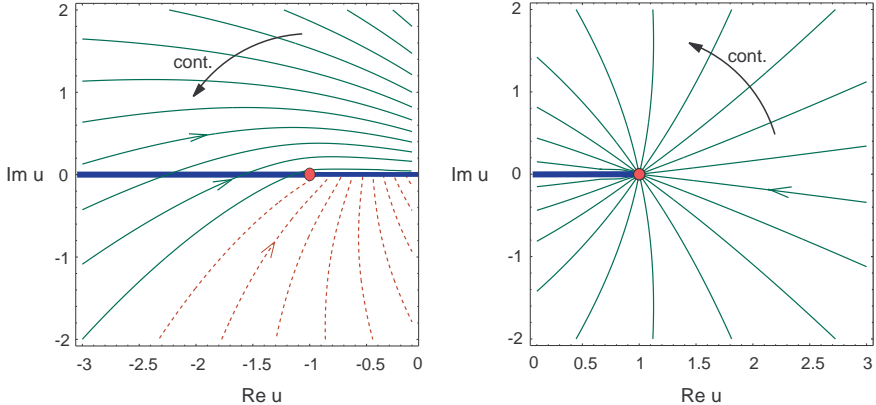,height=5.8cm,angle=0,trim=-40 0 0 0}
  \caption{{\em Right:} Monopole flows in the Seiberg-Witten $u$-plane for $\Re u >
  0$. {\em Left:} continuation of these flows to $\Re u < 0$ by rotating the
  starting point of the flow counterclockwise around $u=\infty$.
  The fat blue lines are period cuts.
  The attractor point is the ``conifold'' point $u=1$, where
  the monopole becomes massless. When performing the monodromy around $u=\infty$, a critical
  flow is encountered (passing through the dyon singularity $u=-1$) beyond which no
  solution exists: the red dashed lines are ``false flows''; they continue beyond the cut
  (not shown on figure) and crash at a regular zero.}
  \label{SWflows}
\end{figure}

Because $Z(u)$ is now analytic, the only possible minima of $|Z|$
are zeros. In fact, it is easy to see from (\ref{atSW}) that the
flows are lines of constant $Z$-phase, which of course necessarily
end on a zero of $Z$. As before, to have a solution, the zero
cannot be at a regular point, so it should be at the singularity
$u=1$ where the monopole becomes massless, or at $u=-1$ where the
elementary dyon becomes massless. Therefore, the only solutions to
(\ref{atSW}) are of the empty hole type: the monopole, with
attractor point $u=1$, and the elementary dyon, with attractor
point $u=-1$, plus of course their oppositely charged partners. In
a neighborhood of their respective attractor points, with the
choice of period cuts shown in fig. \ref{SWflows}, the monopole
has charge $(n_e,n_m)=(0,1)$, while the elementary dyon gets
assigned the charge $(1,1)$ above the cut and $(1,-1)$ below.

Again, we are facing a puzzle. It is well known that at weak
coupling (that is, outside the line of marginal stability given by
$a_D/a \in \IR$), the BPS spectrum also contains a tower of dyons
with $n_m = \pm 1$ and arbitrary (integer) $n_e$. These however
correspond to ``false flows'' breaking down at a regular zero of
$Z=n_e a + n_m a_D$ (on the line of marginal stability). The same
problem arises for the purely electrically charged massive
W-boson.

The paradox can be seen most sharply by starting with a $(0,1)$
monopole flow and performing a $u \to e^{2 i \pi} u$ monodromy
around $u=\infty$ (fig.\ \ref{SWflows}). Doing this monodromy once
should generate a higher dyon with charge $(2,-1)$, doing it twice
should yield a $(-4,1)$ dyon, and so on. However, when circling
around $u=\infty$, at a certain point, one arrives at a critical
flow passing through the $u=-1$ singularity. When one tries to
``pull'' the flow through this singularity, a catastrophe occurs:
due to the nontrivial monodromy of the magnetic charge around
$u=-1$, the flow can no longer end on $u=1$; instead, past the
singularity, it starts to diverge away from the flow just before
criticality, and breaks down at the point (on the line of marginal
stability) where $Z_{(2,-1)}$ (analytically continued along the
flow) becomes zero.

Physically, we don't expect anything really drastic to happen when
we vary the moduli at infinity just a little bit, yet we seem to
find it can cause a complete breakdown of the solution.

So what is going on here?

\sect{Resolutions} \label{resolutions}

\subsection{Composite configurations} \label{splitflow}

\begin{figure}
  \epsfig{file=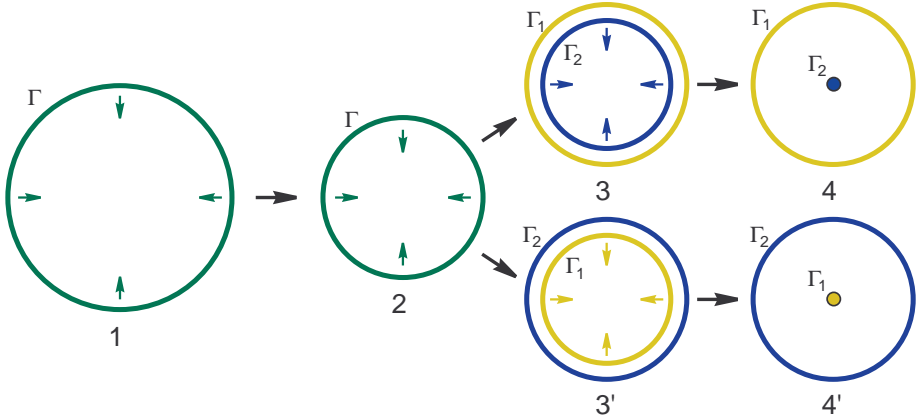,height=6cm,angle=0,trim=-40 0 0 0}
  \caption{The two possible sequences of events for the inmoving shell of
  section \ref{puzzle1}, as described in the text; 1: $\Gamma$-shell contracts, 2:
  marginal stability for decay $\Gamma \to \Gamma_1 + \Gamma_2$ is reached at $r=r_{ms}$ and the
  $\Gamma$-particles split, 3: $\Gamma_1$-shell stays at $r_{ms}$ and $\Gamma_2$ moves further in,
  4: final configuration, a $\Gamma_2$-center surrounded by a $\Gamma_1$-shell, 3',4': same as 3,4
  with $\Gamma_1$ and $\Gamma_2$ interchanged.}
  \label{split}
\end{figure}

We now turn to the resolution of these puzzles. To get a first
hint of what could do the job, consider again the situation
described in section \ref{puzzle1}; the suicidal solution produced
by an inmoving charged shell of charge $\Gamma$. As explained
there, we expect that the modulus at spatial infinity and the
modulus where $Z$ becomes zero are separated by a line of marginal
stability, so we expect the attractor flow to cross this line.
Suppose that this is indeed what happens, say at $r=r_{ms}$, for
the decay process $\Gamma \to \Gamma_1 + \Gamma_2$, and assume for
simplicity that both $\Gamma_1$ and $\Gamma_2$ have well behaved
attractor flows. Then what actually will happen when the shell
shrinks is {\em not} the disaster scenario of section
\ref{puzzle1}; instead, when the particle cloud has reached radius
$R=r_{ms}$, {\em the $\Gamma$-particles will decay into
$\Gamma_1$- and $\Gamma_2$-particles} (fig.\ \ref{split}). Now the
$\Gamma_1$- and $\Gamma_2$-particles cannot {\em both} continue to
move inward, as this would be energetically impossible (the
configuration would no longer be BPS because $Z(\Gamma_1)$ and
$Z(\Gamma_2)$ acquire different phases on points of the
$\Gamma$-flow beyond the marginal stability line). Rather, the
$\Gamma_1$-particles will stay at the marginal stability locus
$r=r_{ms}$, while the $\Gamma_2$-particles move on, or vice versa.
In the first case, when the $\Gamma_2$-charges arrive at $r=0$, we
have a BPS configuration (see below) consisting of a
$\Gamma_2$-charged center surrounded by a $\Gamma_1$-shell at
$r=r_{ms}$. Outside the shell the fields are given by the
$\Gamma$-attractor flow, and inside the shell by the
$\Gamma_1$-attractor flow. In the second case, we have a similar
situation, with 1 and 2 interchanged.

To see that such configurations are indeed BPS, let us compute the
total energy, say for the first case. The energy in the bulk
fields outside the $\Gamma_1$-shell is
$E_{out}=|Z(\Gamma)|_{\infty}-(e^U |Z(\Gamma)|)_{r_{ms}}$. The
energy of the shell itself is $E_{shell}=(e^U
|Z(\Gamma_1)|)_{r_{ms}}$. The energy inside the shell is
$E_{in}=(e^U |Z(\Gamma_2)|)_{r_{ms}}$. So the total energy is
\begin{equation} \label{bpscond}
 E_{tot}=|Z(\Gamma)|_{\infty}+ \left( e^U
 (|Z(\Gamma_1)|+|Z(\Gamma_2)|-|Z(\Gamma_1+\Gamma_2)|) \right)_{r_{ms}} \, .
\end{equation}
But since precisely at marginal stability, the quantity between
brackets is zero, we find indeed $E_{tot}=|Z(\Gamma)|_{r=\infty}$,
that is, the configuration is BPS.

Furthermore, when one would move the shell away from $r=r_{ms}$,
the quantity between brackets becomes strictly positive, so this
configuration is stable under such perturbations.

Another way of seeing this is by considering the force on a test
particle of charge $\epsilon \Gamma_1$ at rest in the attractor
flow field of a charge $\Gamma_2$. This can be derived from
(\ref{source}). As shown in appendix \ref{apA}, the result is that
this force can be derived from the potential
\begin{equation}
 W=2 \, \epsilon \, e^U \, |Z(\Gamma_1)| \,
 \sin^2(\frac{\alpha_1-\alpha_2}{2})\, ,
\end{equation}
where $\alpha_i = \arg Z(\Gamma_i)$. This potential is everywhere
positive, and becomes zero when $\alpha_2=\alpha_1$, that is, at
marginal stability.

It is not difficult to extract the equilibrium radius $r_{ms}$
from the integrated flow equation (\ref{integrated}). Taking the
intersection product of $\Gamma_1$ with this equation gives,
denoting $Z(\Gamma_i)$ in short as $Z_i$:
\begin{equation} \label{imZ1}
 2 \, \Im (e^{-U} e^{-i \alpha} Z_1) = - \langle \Gamma_1,\Gamma \rangle \, \tau
 \, + \, 2 \, \Im (e^{-i \alpha} Z_1)_{\tau=0} \, .
\end{equation}
At $1/\tau=r=r_{ms}$, the left hand side is zero, so
\begin{equation} \label{prermsformula}
 r_{ms} = \frac{\langle \Gamma_1,\Gamma \rangle}{2 \, \Im(e^{-i \alpha} Z_1)_{r=\infty}} \, .
\end{equation}
Using $e^{i \alpha}=Z/|Z|$ with $Z=Z_1+Z_2$ and $\langle
\Gamma_1,\Gamma \rangle = \langle \Gamma_1,\Gamma_2 \rangle$, this
can be written more symmetrically as
\begin{equation} \label{rmsformula}
 r_{ms} = \frac{1}{2} \langle \Gamma_1,\Gamma_2 \rangle \left. \frac{|Z_1+Z_2|}{\Im(\bar{Z_2}
 Z_1)} \right|_{r=\infty}
 \, .
\end{equation}
Some interesting consequences of this identity will be discussed
further on.

Having arrived at this picture of composite configurations in the
approximation of spherical shells, a natural question to ask is
whether supergravity also has nonspherical multicenter BPS
solutions (with nonparallel charges). We will study this problem
in section \ref{compflows}.

\subsection{Forked flows}

\begin{figure}
  \epsfig{file=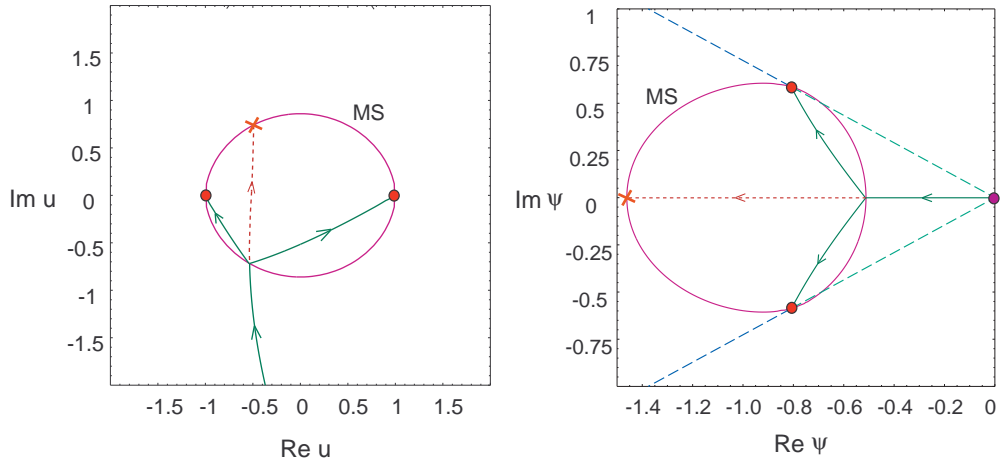,height=6.5cm,angle=0,trim=-20 0 0 0}
  \caption{Numerically computed composite flow for a $(2,-1)$-dyon is Seiberg-Witten
  theory (left) and for
  the state $|10000\rangle_B$ at the Gepner point of the quintic (right). The purple
  ellipsoid line is the relevant line of marginal stability.
  The dotted red line is the false simple attractor flow, crashing on the regular zero of $Z$
  indicated by a red cross. The wedge $4 \pi/5 < \arg \psi < 6 \pi/5$ is indicated by
  dashed lines.}
  \label{compfl}
\end{figure}

The composite configurations discussed above can be represented by
{\em composite flows} (or ``forked flows''): the flow starts as an
ordinary $\Gamma$-attractor flow, reaches a line of marginal
stability, and then splits in a $\Gamma_1$-flow and a
$\Gamma_2$-flow, corresponding to the two possible realization of
the state as a charged center surrounded by a charged shell. The
total energy of the configuration then equals the sum of the
energies associated to each of the constituent flows, that is, for
a $\gamma$-flow running from $i$ to $f$, $E=(e^U |Z(\gamma)|)_f -
(e^U |Z(\gamma)|)_i$.

Thus the generalization to composite spherically symmetric BPS
states simply amounts to the generalization of simple attractor
flows to composite attractor flows. Can we find such composite
flows for the specific examples discussed in sections
\ref{puzzle1} and \ref{puzzle2}? Fortunately, it turns out we can.
As shown in fig.\ \ref{compfl}, the $\Gamma=(2,-1)$ dyon in
Seiberg-Witten theory can be realized as a flow splitting in a
$\Gamma_1=(0,1)$ monopole flow and a $\Gamma_2=2(1,-1)$ elementary
dyon flow. This corresponds, in the supergravity regime with
$\Gamma = N (2,-1)$, $N$ large, to a magnetic core with charge $N
(0,1)$ surrounded by a dyonic shell with charge $2 N (1,-1)$, or
vice versa. The intersection product of an elementary dyon and a
monopole equals $2$.

For the quintic example outlined in section
\ref{puzzle1}, we find a composite flow ending on two copies of
the conifold point (fig.\ \ref{compfl}). In the conventions and
notation of \cite{BDLR}, the state $|10000\rangle_B$ under
consideration has type IIA D-brane charge
$(Q_6,Q_4,Q_2,Q_0)=(2,0,5,0)$, while the charges with vanishing
mass at the two conifold point copies under consideration are
$(-4,-3,-14,10)$ and $(6,3,19,-10)$,\footnote{These slightly
unnaturally looking values arise because the type IIA D-brane
charges are naturally defined only at large volume (or large
complex structure on the IIB mirror). Charges at arbitrary $\psi$
are defined by continuous transport coming from large $\psi$ in
the wedge $0<\arg \psi<2\pi/5$. This procedure assigns charge
$(1,0,0,0)$ to the state with vanishing mass at $\psi=1$. The
states with vanishing mass at the other four copies of the
conifold point get charges related to this one by the $\IZ_5$
monodromy around the Gepner point \cite{BDLR}, which has no reason
to have a particularly nice action when expressed in the type IIA
D-brane basis.} adding up to the required $(2,0,5,0)$. The intersection
product of these two charges equals $5$.

The appearance of these composite flows is very reminiscent of the
appearance of ``3-pronged strings'' in the ``3-1-7 brane picture''
of BPS states in $\N=2$ quantum field theories
\cite{BPS37,threeprong}. This is no coincidence. As explained in
section \ref{staticstring}, in the Seiberg-Witten case for
instance, there is an exact map between the attractor flows and
the stretched strings of \cite{BPS37}. Similarly, the composite
flows arise precisely when the simple geodesic strings fail to
exist and the 3-pronged strings take over, and here again there is
an exact map between the flows and the strings.

Finally note one could imagine more complex configurations,
involving more than one shell, corresponding to more than one flow
split. For now, we will stick to the two charge case however.

\subsection{Monodromy magic}

\begin{figure}
  \epsfig{file=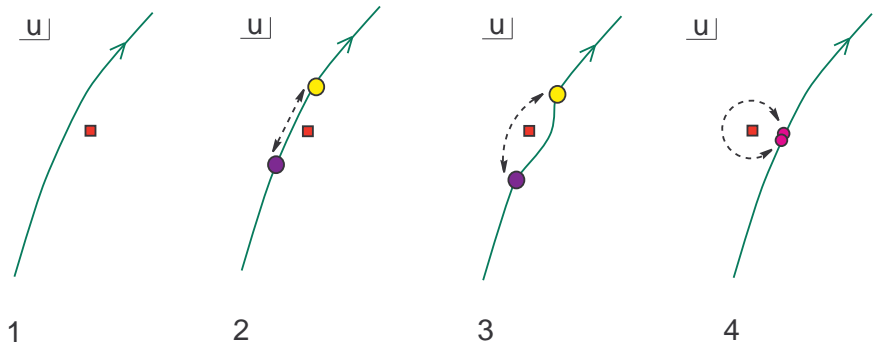,height=5cm,angle=0,trim=-60 0 0 0}
  \caption{1: $u(r)$-flow just before criticality (the red rectangle
  is the point $u=-1$); 2: a virtual monopole-antimonopole pair is created;
  3: flow moves beyond $u=-1$; 4: pair annihilates again but leaves two light
  elementary dyons behind, due to monodromy.}
  \label{dyoncreat}
\end{figure}

This picture also offers a nice way to resolve the monodromy
puzzle of section \ref{puzzle2}. Consider again $N$ monopole
charges at $r=0$, in a vacuum $u_\infty$ such that the attractor
flow is infinitesimally close to the critical one. When we further
vary $u_\infty$ counterclockwise, the flow will pass through the
$u=-1$ point, say at $r=r_c$. Placed at this radius, an elementary
dyon would be massless, so it could be created there at no cost in
energy. And this is precisely what will happen when we continue to
rotate $u_\infty$: $2N$ elementary dyons of charge $(1,-1)$ are
created! Due to the subtleties associated with monodromy, this is
in full agreement with charge conservation. To get some physical
feeling for this phenomenon, suppose we manipulate the
$u(r)$-field in a certain region of space containing a piece of
the surface $r=r_c$, in such way that here the $u(r)$-flow moves
from passing just above to passing just below the $u=-1$
singularity in the $u$-plane (fig.\ \ref{dyoncreat}). Now imagine
that just before the move a virtual monopole-antimonopole pair was
created, to be destroyed again just after the move, and that the
monopole happened to be at $r>r_c$ when the critical trajectory
was crossed, while the antimonopole was at $r<r_c$. Then the
spacetime trajectory of the monopole-antimonopole pair, mapped to
moduli space via $u(r,t)$, encircles the point $u=-1$.
Consequently, there is a monodromy on the monopole charge, of
which the net result is that we are left with two elementary dyons
(with infinitesimal mass) when the monopole-antimonopole pair is
destroyed again! If the resulting configuration is energetically
favorable (so certainly if it is BPS), it will persist. This gives
a physically reasonable mechanism to get the required dyons for
the composite BPS state that is supposed to take over when we
further rotate $u_\infty$. Once the dyon shell is present, we can
continue the monodromy (making the dyon shell massive), till the
(composite) flow passes through the $u=1$ singularity, where the
above process is repeated with massless monopoles.\footnote{The
reader might be puzzled about how our configuration with a dyonic
outer shell gets transformed into one with a magnetic outer shell.
This can be understood from the discussion of empty holes in
section \ref{emptyholes}: when approaching the flow passing
through $u=1$, the distance between the dyonic shell and the
``enhan\c{c}on'' radius $r=r_*$, where $u=1$ is reached and below
which the monopoles cannot be localized, shrinks to zero. Thus, at
the critical flow, the roles of the monopole and dyon shells can
be interchanged continuously.}. In this way we can continue,
creating all expected higher dyons.

Note that a local observer, placed in- or outside the sphere
$r=r_c$, will not note anything peculiar when the transition takes
place. Locally, everything changes perfectly smoothly.

\subsection{Marginal stability, Joyce transitions and
$\Pi$-stability} \label{pistab}

From (\ref{rmsformula}), it follows that when the moduli at
infinity approach the line (or, if the dimension of moduli space
is larger than one, the hypersurface) of marginal stability for
the decay $\Gamma \to \Gamma_1 + \Gamma_2$, the shell radius
$r_{ms}$ will diverge, eventually reaching infinity at marginal
stability. This gives a nicely continuous 4d spacetime picture for
the decay of the state when crossing marginal stability.

Furthermore, (\ref{rmsformula}) tells us at which side of the
marginal stability hypersurface the composite state can actually
exist: since $r_{ms}>0$, it is the side satisfying
\begin{equation} \label{stabcond}
 \langle \Gamma_1,\Gamma_2 \rangle \, \sin(\alpha_1-\alpha_2) > 0 \,
 ,
\end{equation}
where $\alpha_i = \arg Z(\Gamma_i)_{r=\infty}$. Sufficiently close
to marginal stability, this reduces to
\begin{equation}
 \langle \Gamma_1,\Gamma_2 \rangle \, (\alpha_1-\alpha_2) > 0 \, ,
\end{equation}
which is precisely the stability condition for ``bound states'' of
special Lagrangian 3-cycles found in a purely Calabi-Yau
geometrical context by Joyce! (under more specific conditions,
which we will not give here) \cite{joyce,KM}.

Note also that, since the right hand side of (\ref{imZ1}) can only
vanish for one value of $\tau$, the composite configurations we
are considering here will actually satisfy
\begin{equation}
 |\alpha_1-\alpha_2|<\pi \, .
\end{equation}

Another immediate consequence of (\ref{rmsformula}) is the fact
that these composite configurations can only\footnote{at least for
asymptotically flat space. For a space asymptotic to $AdS_2 \times
S^2$, the situation changes.} occur for mutually nonlocal charges,
that is, charges $\Gamma_1$ and $\Gamma_2$ with nonzero
intersection product.

If the constituent $\Gamma_i$ of the composite configuration for
which $\langle \Gamma,\Gamma_i \rangle > 0$ can be identified with
a ``subobject'' of the state as defined in \cite{DFR}, the above
conditions imply that the phases satisfy the $\Pi$-stability
criterion introduced in that reference. Though this similarity is
interesting, it is far from clear how far it extends.
$\Pi$-stability is considerably more subtle than what emerges
here. On the other hand, we have thus far only considered BPS
configurations in a classical, spherical shell approximation, so
also on this side the full stability story can be expected to be
more complicated. We leave this issue for future work.

\sect{The general stationary multicenter case} \label{compflows}

In view of the emergence of composite BPS configuration in the
spherical approximation, it is natural to look for more general
multicenter solutions. This case is far more involved however. In
particular, we have to give up the assumption that the
configuration is static, and allow for more general, but still
stationary, spacetimes.

\subsection{BPS equations}

Stationary (single center) BPS solutions of ${\cal N}=2$
supergravity were first studied in \cite{BLS} from supersymmetry
considerations, in a specific space-dependent $\Omega_0$-gauge
(essentially the one described at the end of section
\ref{bpseom}). Here we will follow an approach based on the
bosonic duality invariant action, similar to the one followed in
section \ref{revisit}, and we let $\Omega_0$ depend on position
only through the the moduli.

Again, we will use the metric ansatz (\ref{stationarymetric}), but
now with $U$ an arbitrary function of position ${\bf x}$, and
$\omega$ not necessarily zero (but still time independent). We
consider only the asymptotically flat case here, that is, $U,
\omega \to 0$ when $r \to \infty$.

We will use boldface notation for 3d quantities as explained in
section \ref{diform}. The 3d Hodge dual with respect to the {\em
flat} metric $\delta_{ij}$ will be denoted by $\s*_0$ , and for
convenience we write $\tomega \equiv e^{2U} \omega$. It will also
turn out to be useful to define the following scalar product of
spatial 2-forms $\sF$ and $\sG$:
\begin{equation}
  ( \sF,\sG ) \equiv \frac{e^{2U}}{1-\tomega^2} \int_X \sF
  \wedge [\, \s*_0 \widehat{\sG} - \s*_0(\tomega \wedge
  \widehat{\sG}) \, \tomega + \s*_0(\tomega \wedge \s*_0 \sG)
  \,] \, .
\end{equation}
Note that we have $( \sF,\sG )=( \sG, \sF )$ and for $\tomega$ not
too large $(\sF,\sF) \geq 0$.

With these assumptions and notations, the action (\ref{S4D}), with
the duality invariant electromagnetic action (\ref{dualinvarS})
substituted in place of the covariant one, becomes, putting
$\gamma \equiv \sqrt{4\pi}$ and dropping a total derivative $\sim
\Delta U$ from the gravitational action:
\begin{eqnarray}
 S_{4D}=- \frac{1}{16 \pi} \int dt \int_{\IR^3} && \!\!\!\!\!\! \{ \, 2 \, \sd U \wedge \s*_0 \sd U
 \, - \frac{1}{2} e^{4U} \sd \omega \wedge \s*_0 \sd \omega  \nonumber \\
 && \!\!\!\! + 2 g_{a\bb} \, \sd z^a \wedge \s*_0 \sd \bz^{\bb}
  \, + \, \left( \sF , \sF \right) \} \, .
  \label{stationaryaction}
\end{eqnarray}
We will derive the BPS equation by ``squaring'' the action in a
way inspired by (\ref{square2}). Let $\alpha$ be an arbitrary real
function on $\IR^3$, denote
\begin{equation}
 \sD \equiv \sd + i \, ( \sQ + \sd\alpha + \frac{1}{2} e^{2U} \s*_0
 \sd \omega) \, , \label{sDdef}
\end{equation}
with $\sQ$ as in (\ref{Qdef}), and define the 2-form $\sG$ as
\begin{equation}
\sG \equiv \sF - 2 \, \Im \s*_0 \sD(e^{-U} e^{-i\alpha} \Omega)
 + 2 \, \Re \, \sD(e^{U} e^{-i\alpha} \Omega \, \omega) \, ,
 \label{sGdef}
\end{equation}
Then we find for the integrand ${\cal L}$ of
(\ref{stationaryaction}), after some calculational effort
involving repeated use of the identities (\ref{hodgediag}) and
(\ref{calculus1})-(\ref{calculus3}),
\begin{eqnarray}
 {\cal L} &=& (\sG,\sG)
 -\, 4 \, (\sQ + \sd\alpha + \frac{1}{2} e^{2U} \s*_0 \sd\omega)
 \wedge \Im \langle \sG, e^U e^{-i\alpha} \Omega \rangle  \nonumber \\
 && + \, \sd \, [\, 2 \, \tilde{\omega} \wedge (\sQ+ \sd\alpha) + 4 \, \Re \langle \sF, e^U e^{-i\alpha}
 \Omega \rangle \,] \, .
\end{eqnarray}
Thus if
\begin{eqnarray}
 \sG &=& 0 \label{bpsmc1} \\
 \sQ + \sd\alpha + \frac{1}{2} e^{2U} \s*_0 \sd\omega &=& 0 \, ,
 \label{bpsmc2}
\end{eqnarray}
we have a BPS solution to the equations of motion following from
the reduced action (\ref{stationaryaction}) \footnote{We will
assume that these solutions also satisfy the equations of motion
of the full action without restrictions on the metric, as in the
spherically symmetric case, though we did not check this
explicitly.} (we will verify the saturation of the BPS bound
below). Now from (\ref{bpsmc2}) and (\ref{sDdef}), we have $\sD =
\sd$, and (\ref{bpsmc1}) becomes
\begin{equation}
 \sF + 2 \, \sd \, \Re (e^{U} e^{-i\alpha} \Omega \, \omega)
 = 2 \s*_0 \sd \, \Im (e^{-U} e^{-i\alpha} \Omega)
 \, .
 \label{bpsmc3}
\end{equation}
Since by construction $\sd \sF=0$ (away from sources), this
implies $\sd \s*_0 \sd \, \Im (e^{-U} e^{-i\alpha} \Omega) = 0$,
so we can write
\begin{equation} \label{bpsmc4}
 2 \, \Im (e^{-U} e^{-i\alpha} \Omega) = H \, ,
\end{equation}
with $H$ a $H^3(X,\IZ)$-valued harmonic function on $\IR^3$
(possibly with source singularities). If we take the sources to be
at positions $\sx_i$ with charges $\Gamma_i$, where
$i=1,\ldots,N$, then from (\ref{bpsmc3}) and (\ref{flux}), we
obtain
\begin{equation} \label{Hform}
 H = -\sum_{i=1}^N \Gamma_i \, \tau_i \, + \, 2 \, \Im (e^{-i \alpha}
 \Omega)_{r=\infty} \, ,
\end{equation}
with $\tau_i=1/|\sx-\sx_i|$. Defining the 1-form
\begin{equation}
 \szeta \equiv - \langle \sd H,\Omega \rangle = \sum_{i=1}^N Z(\Gamma_i) \, \sd \tau_i \, ,
\end{equation}
we get from taking intersection products of $\sd H$ given by
(\ref{bpsmc4}) with $\Omega$ and $D_a \Omega$, and using
(\ref{calculus1})-(\ref{calculus3}):
\begin{eqnarray}
 \sQ + \sd \alpha &=& e^U \Im (e^{-i\alpha} \szeta) = -\frac{1}{2} e^{2U}
 \langle \sd H, H \rangle \label{bpsmc5} \\
 \sd U &=& - e^U \Re (e^{-i\alpha} \szeta) \label{bpsmc6} \\
 \sd z^a &=& - e^U g^{a\bb} e^{i\alpha} {\bar{D}}_{\bar{b}} \szeta
 \, . \label{bpsmc7}
\end{eqnarray}
Using (\ref{bpsmc5}), equation (\ref{bpsmc2}) can be rewritten as:
\begin{equation} \label{omeq}
 \s*_0 \sd \omega = \langle \sd H, H \rangle \, .
\end{equation}
Equations (\ref{bpsmc4}) and (\ref{Hform}) generalize
(\ref{integrated}). Given the sources and the moduli at infinity,
they yield the fields $U(\sx)$, $\alpha(\sx)$ and $z^a(\sx)$.
Equation (\ref{omeq}) on the other hand gives $\omega(\sx)$ (up to
gauge transformations $\omega \to \omega + \sd f$, which can be
absorbed by a coordinate transformation $t \to t - f$). Equations
(\ref{bpsmc6}) and (\ref{bpsmc7}) generalize the flow equations
(\ref{at1})-(\ref{at2}).

Note that asymptotically for $1/\tau = r \to \infty$, the right
hand side of (\ref{bpsmc5}) vanishes and $\zeta \to \sum_i
Z(\Gamma_i) \sd \tau$, implying
\begin{equation}
 \alpha \to \arg Z(\Gamma) \quad \mbox{ and } \quad \zeta \to Z(\Gamma) \sd \tau
  \quad \quad \mbox{when } r \to
 \infty \, ,
\end{equation}
where $\Gamma=\sum_i \Gamma_i$. Thus, far from all sources, we
have again a simple attractor flow, corresponding to the total
charge $\Gamma$, as could be expected physically. In particular
(\ref{bpsmc6}) gives $\sd U \to - |Z(\Gamma)| d\tau$, with
$\tau=1/r$, establishing the saturation of the BPS bound on the
mass:
\begin{equation}
 M_{ADM} = |Z(\Gamma)|_{r=\infty} \, .
\end{equation}
In the spherically symmetric case (and in the multicenter case
with parallel charges), the above asymptotics become exact, and we
retrieve the equations found earlier for those cases. Similarly,
close to the center $\sx_i$, we have
\begin{equation}
 \alpha \to \arg Z(\Gamma_i) \quad \mbox{ and } \quad \zeta \to Z(\Gamma_i) \sd
 \tau_i
 \quad \quad \mbox{when } \sx \to \sx_i \,
 ,
\end{equation}
and again we have asymptotically the flow equations for a single
charge attractor, as could be expected physically. In particular
the moduli at $\sx_i$ will be fixed at the $\Gamma_i$-attractor
point.

The BPS equations of motion for the moduli and the metric obtained
here can be seen to reduce to the equations found in \cite{BLS} in
the $\Omega_0$-gauge described at the end of section \ref{bpseom},
except that we do not find the restricition $\sd \sQ=0$.

\subsection{Some properties of solutions}

Consider a multicenter solution, with distinct centers $\sx_i$,
$i=1,\ldots,n$, to the BPS equations
\begin{eqnarray}
 2 \, e^{-U} \Im (e^{-i\alpha} \Omega) &=& H \, , \label{mc1} \\
 \s*_0 \sd \omega &=& \langle \sd H, H \rangle \, , \label{mc2}
\end{eqnarray}
where
\begin{equation}
 H = -\sum_{i=1}^n \Gamma_i \, \tau_i \, + \, 2 \, \Im (e^{-i
 \alpha}
 \Omega)_{r=\infty} \, , \label{Haha}
\end{equation}
as derived in the previous section. Acting with $\sd \s*_0$ on
equation (\ref{mc2}) gives
\begin{equation}
  0 = \langle \Delta H, H \rangle \, ,
\end{equation}
so, using (\ref{Haha}) and $\Delta \tau_i = -4 \pi
\delta^3(\sx-\sx_i)$, we find that for all $i=1,\ldots,n$:
\begin{equation} \label{distconstr}
  \sum_{j=1}^n \frac{\langle \Gamma_i,\Gamma_j
  \rangle}{|\sx_i-\sx_j|} = 2 \, \Im(e^{-i \alpha}
  Z(\Gamma_i))_{\infty} \, .
\end{equation}
In the particular case of one source with charge $\Gamma_2$ at
$\sx=0$ and $m$ sources with equal charge $\Gamma_1$ at positions
$\sx_i$, this becomes
\begin{equation}
 |\sx_i| = \frac{\langle \Gamma_1,\Gamma_2
 \rangle}{2 \, \Im(e^{-i \alpha} Z(\Gamma_1))_{\infty}} \, ,
\end{equation}
which is equal to the equilibrium distance $r_{ms}$ found in the
spherical shell picture, equation (\ref{prermsformula}).

In general, the moduli space of solutions to (\ref{distconstr}) will be
quite nontrivial. Some general properties can be deduced relatively easy however.
For instance, in a configuration made of only two different charge types
$\Gamma_1$ and $\Gamma_2$ (distributed over an arbitrary number of centers), the
charges of different type, if mutually nonlocal,
will be driven to infinite distance from each other when
$(\Gamma_1,\Gamma_2)$-marginal stability is approached. This is similar to what we found
for the spherical shell case. The stability condition (\ref{stabcond}) reappears
as well. If on the other hand
the two charges are mutually local (zero intersection),
no BPS configuration exists with the two charges separated from each other, unless
their phases are equal, that is, at marginal stability (then we can place the charges
anywhere).\footnote{If we consider a spacetime asymptotic to $AdS_2 \times S^2$
instead of the asymptotically flat one we are assuming here, mutually local charges
are no longer constrained by (\ref{distconstr}), because
there will be an additional factor $\exp[-U(r=\infty)] \equiv 0$ on the right hand side
of (\ref{distconstr}).}

For configurations made of more charge types, things
get more complicated, but we will not go into this here.

\begin{figure}
  \epsfig{file=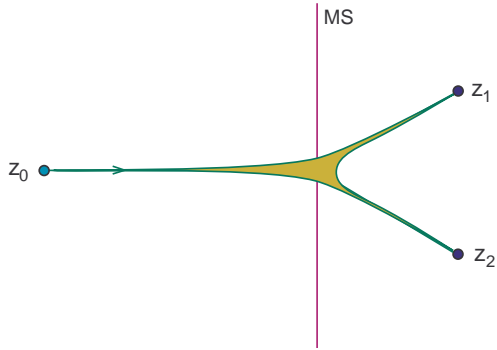,height=5cm,angle=0,trim=-150 0 0 0}
  \caption{Sketch of the image of $z(\sx)$ in moduli space for a
  multicenter solution containing two different charges $\Gamma_1$
  and $\Gamma_2$, with attractor
  points $z_1$ resp.\ $z_2$, and modulus at spatial infinity $z_0$. The line
  labeled ``MS'' is a $(\Gamma_1,\Gamma_2)$-marginal stability line.}
  \label{fattened}
\end{figure}

Finally, note that because of the asymptotics discussed at the end
of the previous section, we can expect the image of the moduli
fields into moduli space for a multicenter configuration with only
two different charge types to look like a fattened version of the
composite flows we introduced earlier to represent the composite
spherical shell configurations (fig. \ref{fattened}). Furthermore,
we can expect that the more spherically symmetric the multicenter
configuration becomes, the more this fattened version will
approach the one dimensional composite flow. This is similar to
what was found for spatial descriptions of dyons in $\N=4$
(effective) quantum field theories \cite{n4,KK}.

\subsection{Angular momentum}

It is well known from ordinary Maxwell electrodynamics that multicenter
configurations with mutually nonlocal charges (e.g.\ the monopole-electron system) can
have intrinsic angular momentum even when the particles are at rest. The same
turns out to be true here.

We define the angular momentum vector $\sJ$ from the asymptotic
form of the metric (more precisely of $\omega$) as \cite{MTW}
\begin{equation}
  \omega_i = 2 \, \epsilon_{ijk} \, J^j \, \frac{x^k}{r^3} \, + O(\frac{1}{r^3})
  \quad \quad \mbox{for } r \to \infty \, .
\end{equation}
Plugging this expression in (\ref{mc2}) and using (\ref{Haha}) and (\ref{distconstr}),
we find
\begin{equation}
  \sJ = \frac{1}{2} \sum_{i<j} \langle \Gamma_i, \Gamma_j \rangle \, {\se}_{ij} \, ,
\end{equation}
where ${\se}_{ij}$ is the unit vector pointing from $\sx_j$ to $\sx_i$:
\begin{equation}
  {\se}_{ij} = \frac{\sx_i - \sx_j}{|\sx_i - \sx_j|} \, .
\end{equation}
Just like in ordinary electrodynamics, this is a ``topological'' quantity: it is
independent of the details of the solution and quantized in half-integer units (more
precisely, when all charges are on the z-axis, $2 J_z \in \IZ$).

The appearance of intrinsic configurational
angular momentum implies that quantization of these
composites will have some nontrivial features.

\sect{Conclusions}

We have shown the emergence of some puzzles and paradoxes arising
when one tries to construct four dimensional low energy effective
supergravity solutions corresponding to certain BPS states in type
II string theory compactified on a Calabi-Yau manifold, and
demonstrated how these can be resolved by considering composite
and extended configurations. We made connections to the
enhan\c{c}on mechanism, the 3-pronged string picture of QFT BPS
states, $\Pi$-stability and Joyce transitions of special
Lagrangian manifolds. The problem was analyzed in a spherical
shell approximation and by considering multicenter BPS solutions.

There are quite some problems however, new and old ones, we didn't
touch upon. The most prominent one is that we didn't analyze to
what extent these states really exist as BPS bound states in the
full quantum theory. It seems quite likely that we now face the
opposite problem we started with: instead of too little, we might
now have too many possible solutions. In view of the nontriviality
of quantum mechanics with mutually nonlocal charges, it is not
unconceivable that a proper semiclassical treatment would
eliminate some of these spurious solutions. But even at the
classical level the existence issue is not completely settled. We
did not show for instance that all solutions to (\ref{distconstr})
actually lead to well-behaved BPS solutions to the equations of
motion; the same phenomenon causing the breakdown of some naively
expected spherically symmetric solutions, namely hitting a zero of
the central charge, could cause naively expected solutions to
break down in this more complicated setting as well.

In this setup it seems also quite possible that a certain charge
can have several different realizations as a BPS solution in a
given vacuum, for example both as a single center and as a two
center configuration. Crossing a line of marginal stability could
then cause one realization to disappear, while leaving the other
intact. The D-brane analog of this would presumably be a ``jump''
in its moduli space. This brings us to another interesting open
question: is there a connection between D-brane moduli spaces and
supergravity solution moduli spaces? And could those solution
moduli spaces (for asymptotically flat or $AdS_2 \times S^2$
spacetimes) teach us something about black hole entropy?

It could also be worthwhile to further explore the relation with
$\Pi$-stability, briefly mentioned in section \ref{pistab}.

Finally, this and other recent work \cite{enhancon,myers}
illustrates an apparently recurrent theme in string theory: the
resolution of singularities by creation of finitely extended
D-brane configurations. It would be interesting to find out what
the dielectric, non-commutative D-brane effects of \cite{myers}
can teach us about the states described in this paper.

\vskip 10mm \noindent {\bf \large Acknowledgements}

\vskip 5mm \noindent

I would like to thank Michael Douglas, Brian Greene, Calin
Lazaroiu, Gregory Moore, Robert Myers, Christian R\"omelsberger,
Walter Troost and Eric Zaslow for useful discussions and
correspondence.

\appendix

\sect{Potential for a test charge} \label{apA}

From (\ref{source}), it follows that the Lagrangian (with respect
to the time coordinate $t$) for a test charge $\Gamma_t$ at rest
in the attractor flow field of a charge $\Gamma$ is (denoting
$Z(\Gamma_t)$ in short as $Z_t$, and similarly for the other
quantities involved)
\begin{equation}
 L = - e^U |Z_t| - \frac{\sqrt{\pi}}{\gamma} \langle \Gamma_t,
 {\cal A}_0 \rangle \,
\end{equation}
where ${\cal A}_0$ is obtained from (\ref{fullemfield}):
\begin{equation}
 \partial_i {\cal A}_0 = \frac{\gamma}{\sqrt{4 \pi}} \, e^{2U}
 \partial_i \tau \, \widehat{\Gamma} \, .
\end{equation}
From (\ref{bps3}), we get
\begin{equation}
  \Gamma = i\, \partial_\tau (e^{-U} \, e^{-i \alpha} \Omega)
  \, + \, \mbox{c.c.} \, ,
\end{equation}
so, using (\ref{calculus3}) and (as shown in section \ref{bpseom})
$Q_\tau + \dot{\alpha} = 0$:
\begin{equation}
  \Gamma = - i\, e^{-U} \, \dot{U} \, e^{-i \alpha} \Omega
  + i\, e^{-U} e^{-i \alpha} D_a \Omega \, \dot{z}^a
  \, + \, \mbox{c.c.} \, ,
\end{equation}
hence from (\ref{hodgediag}) and again (\ref{calculus3}):
\begin{eqnarray}
 \widehat{\Gamma} &=& - e^{-U} \, \dot{U} \, e^{-i \alpha} \Omega
  - e^{-U} e^{-i \alpha} D_a \Omega \, \dot{z}^a
  \, + \, \mbox{c.c.} \\
  &=& - e^{-2 U} \partial_\tau (e^U \, e^{-i \alpha} \Omega) \, + \,
  \mbox{c.c.} \, .
\end{eqnarray}
Therefore
\begin{eqnarray}
 \partial_i(\frac{\sqrt{\pi}}{\gamma} \langle \Gamma_t,
 {\cal A}_0 \rangle) &=& \frac{1}{2} e^{2U} \, \partial_i \tau \,
 \langle \Gamma_t,\widehat{\Gamma} \rangle \\
 &=& - \partial_i \left( e^U \Re(e^{-i \alpha} Z_t) \right) \, ,
\end{eqnarray}
and thus (up to a constant)
\begin{eqnarray}
  L &=& - e^U |Z_t| + e^U \Re(e^{-i \alpha} Z_t)  \\
  &=& - e^U |Z_t| \left( 1- \cos(\alpha_t-\alpha) \right) \\
  &=& - 2 e^U |Z_t| \sin^2(\frac{\alpha_t-\alpha}{2}) \, .
  \nonumber
\end{eqnarray}
The force on the test particle is $F_i = \partial_i L$, so we find
for the force potential, as announced in section \ref{splitflow}:
\begin{equation}
 W = 2 e^U |Z_t| \sin^2(\frac{\alpha_t-\alpha}{2}) \, .
\end{equation}

\end{document}